\documentclass[acmtog,authorversion]{acmart}

\AtBeginDocument{%
  }

\graphicspath{{./images/}}
\usepackage{braket}

\definecolor{dblue}{rgb}{0.0,0.0,0.5}
\definecolor{dgreen}{rgb}{0.0,0.5,0.0}
\definecolor{dred}{rgb}{0.6,0.0,0.0}
\definecolor{dorange}{rgb}{0.6,0.25,0.0}
\definecolor{dyellow}{rgb}{0.5,0.5,0.0}

\newcommand{\ignorethis}[1]{}

\renewcommand{\textrightarrow}{$\rightarrow$}

\def\eg{{e.g.}}
\def\ie{{i.e.}}
\def\etal{{et al.}}

\providecommand{\DIFdel}[1]{}
\renewcommand{\DIFdel}[1]{}

\providecommand{\DIFadd}[1]{}
\renewcommand{\DIFadd}[1]{\textcolor{blue}{#1}}

\copyrightyear{2025}
\acmYear{2025}
\setcopyright{cc}
\setcctype{by}
\acmConference[SIGGRAPH Conference Papers '25]{Special Interest Group on Computer Graphics and Interactive Techniques Conference Conference Papers }{August 10--14, 2025}{Vancouver, BC, Canada}
\acmBooktitle{Special Interest Group on Computer Graphics and Interactive Techniques Conference Conference Papers (SIGGRAPH Conference Papers '25), August 10--14, 2025, Vancouver, BC, Canada}
\acmDOI{10.1145/3721238.3730595}
\acmISBN{979-8-4007-1540-2/2025/08}

\author{Chong Zeng}
\affiliation{%
  \institution{State Key Lab of CAD \& CG, Zhejiang University and Microsoft Research Asia}
  \city{Hangzhou}
  \country{China}
}
\orcid{0009-0004-6373-6848}
\email{chongzeng2000@gmail.com}

\author{Yue Dong}
\affiliation{%
  \institution{Microsoft Research Asia}
  \city{Beijing}
  \country{China}
}
\orcid{0000-0003-0362-337X}
\email{yuedong@microsoft.com}

\author{Pieter Peers}
\affiliation{%
  \institution{College of William \& Mary}
  \city{Williamsburg}
  \country{USA}
}
\orcid{0000-0001-7621-9808}
\email{ppeers@siggraph.org}

\author{Hongzhi Wu}
\affiliation{%
  \institution{State Key Lab of CAD \& CG, Zhejiang University}
  \city{Hangzhou}
  \country{China}
}
\orcid{0000-0002-4404-2275}
\email{hwu@acm.org}

\author{Xin Tong}
\affiliation{%
  \institution{Microsoft Research Asia}
  \city{Beijing}
  \country{China}
}
\orcid{0000-0001-8788-2453}
\email{xtong@microsoft.com}

\renewcommand{\shortauthors}{Zeng et al.}

\begin{document}

\title{RenderFormer: Transformer-based Neural Rendering of Triangle Meshes with Global Illumination}

\renewcommand{\shortauthors}{}
\citestyle{acmauthoryear}

\begin{abstract}
    
  We present RenderFormer, a neural rendering pipeline that directly
  renders an image from a triangle-based representation of a scene
  with full global illumination effects and that does not require
  per-scene training or fine-tuning.  Instead of taking a
  physics-centric approach to rendering, we formulate rendering as a
  sequence-to-sequence transformation where a sequence of tokens
  representing triangles with reflectance properties is converted to a
  sequence of output tokens representing small patches of
  pixels. RenderFormer follows a two stage pipeline: a
  view-independent stage that models triangle-to-triangle light
  transport, and a view-dependent stage that transforms a token
  representing a bundle of rays to the corresponding pixel values
  guided by the triangle-sequence from the view-independent stage.
  Both stages are based on the transformer architecture and are
  learned with minimal prior constraints.  We demonstrate and evaluate
  RenderFormer on scenes with varying complexity in shape and light
  transport.
\end{abstract}

  


\newcommand{\teaserFigWidth}{0.25\textwidth}

\begin{teaserfigure}
  \centering
\addtolength{\tabcolsep}{-5.5pt}
 \begin{tabular}{cccc}
    \includegraphics[width=\teaserFigWidth]{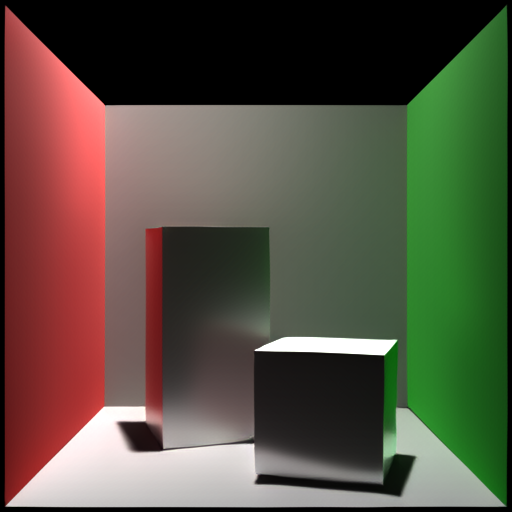}
    &\includegraphics[width=\teaserFigWidth]{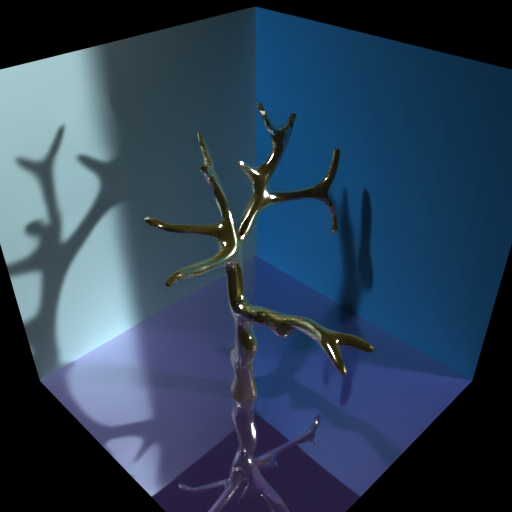} &
    \includegraphics[width=\teaserFigWidth]{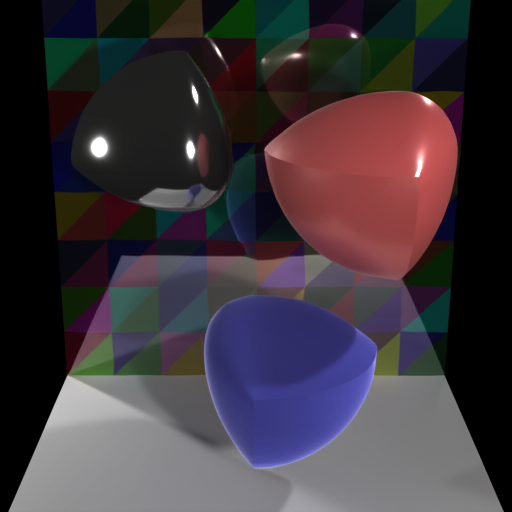}
    &\includegraphics[width=\teaserFigWidth]{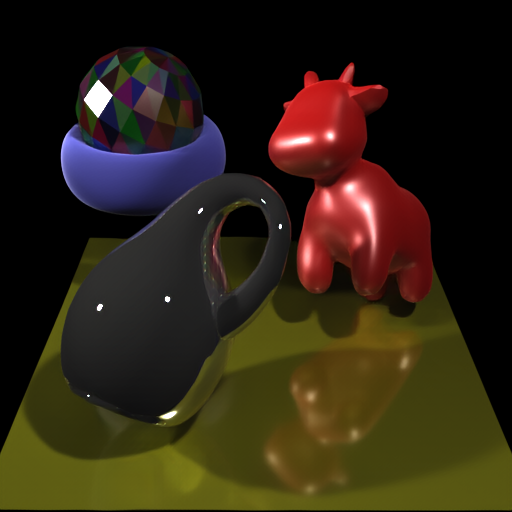} \\
 \end{tabular}
 \caption{Examples of triangle-mesh based scenes rendered with
   RenderFormer without per-scene training or fine-tuning that include
   (multiple) specular reflections, complex shadows with details finer
   than a triangle, diffuse indirect lighting, glossy reflections,
   soft and hard shadows, and multiple light sources.}
 \label{fig:teaser}
\end{teaserfigure}

\maketitle

\section{Introduction}

Traditional graphics pipelines render virtual scenes by simulating the
physical process of light transport through the scene.  Recently,
neural rendering has endeavored to bypass the simulation process, and
instead \emph{learn} to predict the effects of light
transport. However, most neural rendering methods often accomplishes
this by overfitting the model to a fixed scene. This raises the
intriguing question whether it is possible to learn a rendering
\emph{pipeline} rather than a rendering \emph{model}.

In this paper, we take a first step towards a \emph{fully} neural
rendering pipeline, named RenderFormer, that (a) does not require
per-scene training, (b) takes a classic triangle-mesh based scene
description as input, and (c) that renders the scene with full global
illumination. To achieve these goals, we offer a new perspective on
resolving light transport in a virtual scene, and formulate rendering
as a sequence-to-sequence transformation, where each token in the
sequence represents a triangle with reflectance properties that is
subsequently transformed to a triangle with the converged radiance
distribution of the light transport equilibrium. Rather than
explicitly describing the resulting radiance distribution and
following the flow of light through the virtual scene as dictated by
the Rendering Equation~\cite{Kajiya:1986:RE}, RenderFormer learns a
neural rendering pipeline directly from data with minimal prior
constraints.  Compared to conventional rendering paradigms:
RenderFormer directly ‘solves’ the rendering equation without
Monte-Carlo integration noise and without requiring complex
algorithmic modification as in rasterization. In contrast to existing
neural rendering methods for synthetic scenes, RenderFormer does not
require per-scene/object training (e.g., the objects
in~\autoref{fig:teaser} are not part of the training set).

RenderFormer follows a two-stage architecture: a
view-in\-de\-pen\-dent stage that models triangle-to-triangle light
transport, and a view-dependent stage that evaluates the transformed
sequence of triangle tokens into an image.  Both stages are based on
the powerful transformer architecture~\cite{Vaswani:2017:AIA} known
for its capability to model long-range relations (\eg, light transport
from one triangle to all other triangles).  However, in contrast to
typical transformer architectures, RenderFormer utilizes a (relative)
positional encoding based on the $3$D spatial position of the
triangles rather than the $1$D index position in the sequence.
Similar to most neural rendering techniques, RenderFormer does not
require recursive computations and directly solves global illumination
transport in a single pass.  Moreover, RenderFormer is fully based on
learnable neural components, and thus naturally fully differentiable,
without relying on existing fixed (i.e., non-learnable) rendering
algorithms such as rasterization, ray tracing, or ray marching.

In this paper, we present an initial step towards a full neural
rendering pipeline on a constrained set of scene types.  First, due to
the computational costs of transformers, RenderFormer is currently
limited to triangle meshes of at most $4,\!096$ triangles.  Second,
RenderFormer is also constrained by the variations seen during
training: currently our training data only includes a single
reflectance model~\cite{Walter:2007:MMR}, with its parameters assigned
on a per-triangle basis (\ie, no textures). The training scenes
include at most $8$ diffuse light sources, and the camera (with fixed
$512 \times 512$ resolution) is placed outside the scene's bounding
box.  We believe that RenderFormer, with further development and
optimization, can potentially offer an alternative rendering paradigm
for both forward and inverse rendering applications while leveraging
current (and future) advances in transformer-optimized tensor cores.

\section{Related Work}
\label{sec:related}

\paragraph{Rendering Equation}
The Rendering Equation~\cite{Kajiya:1986:RE} formally describes light
transport in virtual scenes defined by its geometry (often modeled by
a triangle mesh), the associated material properties in the form of
Bidirectional Reflectance Distribution Functions
(BRDFs)~\cite{Nicodemus:1992:GCN}, light sources, and a virtual
camera.  Over the past four decades a rich variety of methods have
been proposed for accurately and efficiently solving the Rendering
Equation.  Monte Carlo path tracing and
variants~\cite{Dutre:2018:AGI,Pharr:2023:PBR} are among the most
popular and effective methods for solving the rendering
equation. Recently, to offset the significant computational cost, path
tracing algorithms have been augmented by machine learning
techniques~\cite{Wang:2023:SAD} (\eg, filtering~\cite{Bako:2017:KPC}
or caching~\cite{Muller:2021:RTN,Coomans:2024:RTN}).  All of the above
methods explicitly encode the rendering equation as part of the
solution method, and hence due to the recursive nature of the
Rendering Equation, these methods are also recursive.  RenderFormer
does not rely explicitly on the Rendering Equation, and directly
computes light transport without recursion.

An alternative class of mathematical techniques for solving the
Rendering Equation are finite element methods, and the resulting class
of rendering algorithms are called \emph{radiosity}
methods~\cite{Goral:1984:MTI,Cohen:1988:PRA}. However, classic
radiosity methods are mostly limited to isotropic scattering from
diffuse surfaces.  In the spirit of radiosity,
Hadadan~\etal~\shortcite{Hadadan:2021:NR} introduce a
learning-inspired \emph{neural radiosity} variant that decouples
solving the Rendering Equation (\ie, training) and rendering (\ie,
inference) that can effectively synthesize arbitrary views of a scene
without material restrictions.  Neural radiosity (and its extension to
dynamic scenes~\cite{Su:2024:DNR}) shares similarities to Precomputed
Radiance Transfer (PRT)~\cite{Sloan:2023:PRT} that formulates light
transport as an inner-product between the light transport matrix of a
scene and the lighting expressed in a suitable basis (\eg, Spherical
Harmonics).  Inspired by PRT, Rainer~\etal~\shortcite{Rainer:2022:NPR}
and Gao~\etal~\shortcite{Gao:2022:NGI} leverage neural networks to
learn a suitable embedding of the incident lighting instead of a
predefined lighting basis.  Neural radiosity and PRT methods incur a
large precomputation overhead for \emph{each} scene.  While training
RenderFormer is expensive, training only happens once, after which
scenes can be rendered without further training.

\paragraph{Neural Rendering}
Neural rendering methods~\cite{Tewari:2022:ANR} replace the simulation
of light transport by a learned neural process, and hence the
Rendering Equation is never explicitly imposed and instead an
implicit representation of light transport is learned, Neural Radiance
Fields (NeRFs)~\cite{Mildenhall:2021:NRS} is a well known example of
such an implicit representation of light transport.  Neural rendering
in general employs a neural scene representation which requires
specialized
methods~\cite{Granskog:2020:CNS,Granskog:2021:NSG,Yuan:2022:NGE,Haque:2023:INE,Zheng:2024:NGI}
for making scene modifications. In contrast, RenderFormer takes a
regular triangle-mesh based scene description as input, and thus is
compatible with existing tool-chains for authoring virtual scenes.

Sanzenbacher~\etal~\shortcite{Sanzenbacher:2020:LNL} perform
screen-space neural shading augmented with global neural light
transport computed on a point-cloud representation of the scene. While
the point-cloud helps to generalize the light transport computations,
the two stage network is trained per scene (either static or dynamic).
In contrast RenderFormer does not require per scene training.
RenderNet~\cite{Nguyen:2018:RAD} and Neural Voxel
Rendering~\cite{Rematas:2020:NVR} learn a convolutional neural
rendering pipeline. However, instead of triangle meshes, both methods
take a 3D voxel grid as input, and only learn \emph{local} shading
under a single point light. In contrast, RenderFormer renders the
scene with full global illumination.

\paragraph{Transformers for Rendering}
The key building block in RenderFormer is the transformer
architecture~\cite{Vaswani:2017:AIA} which is built around multi-head
attention blocks and that maps a sequence of tokens to another
sequence of tokens while handling long-range
dependencies. Transformers have proven to be effective architectures
for vision tasks~\cite{Dosovitskiy:2020:IWW} and for driving large
language models~\cite{Kenton:2019:BPT}.  In rendering,
Ren~\etal~\shortcite{Ren:2024:LLO} leverage the cross-attention
mechanism in transformers for accelerating the gather step in neural
reflective shadow maps. In the context of NeRFs,
NerFormer~\cite{Reizenstein:2021:COI} leverages epipolar constraints
and attention to construct feature
volumes. IBRNet~\cite{Wang:2021:ILM} estimates density along rays with
transformers. Recent view-interpolation
methods~\cite{Sajjadi:2022:SRT,Varma:2022:IAA,Suhail:2022:LFN,Liang:2024:RMR},
not only employ transformers to compute features along rays, but also
use a transformer to aggregate features along rays.
LVMS~\cite{Jin:2024:LAL} takes a different approach, and directly
transforms pixel patches from the input images to view-interpolated
images. Jin~\etal~ encode the poses of the input and output cameras by
tokenizing each pixels' view ray; RenderFormer uses a similar strategy
for tokenizing the pose of the virtual camera.


\section{RenderFormer}
\label{sec:renderformer}

\begin{figure*}
  \includegraphics[width=0.95\textwidth]{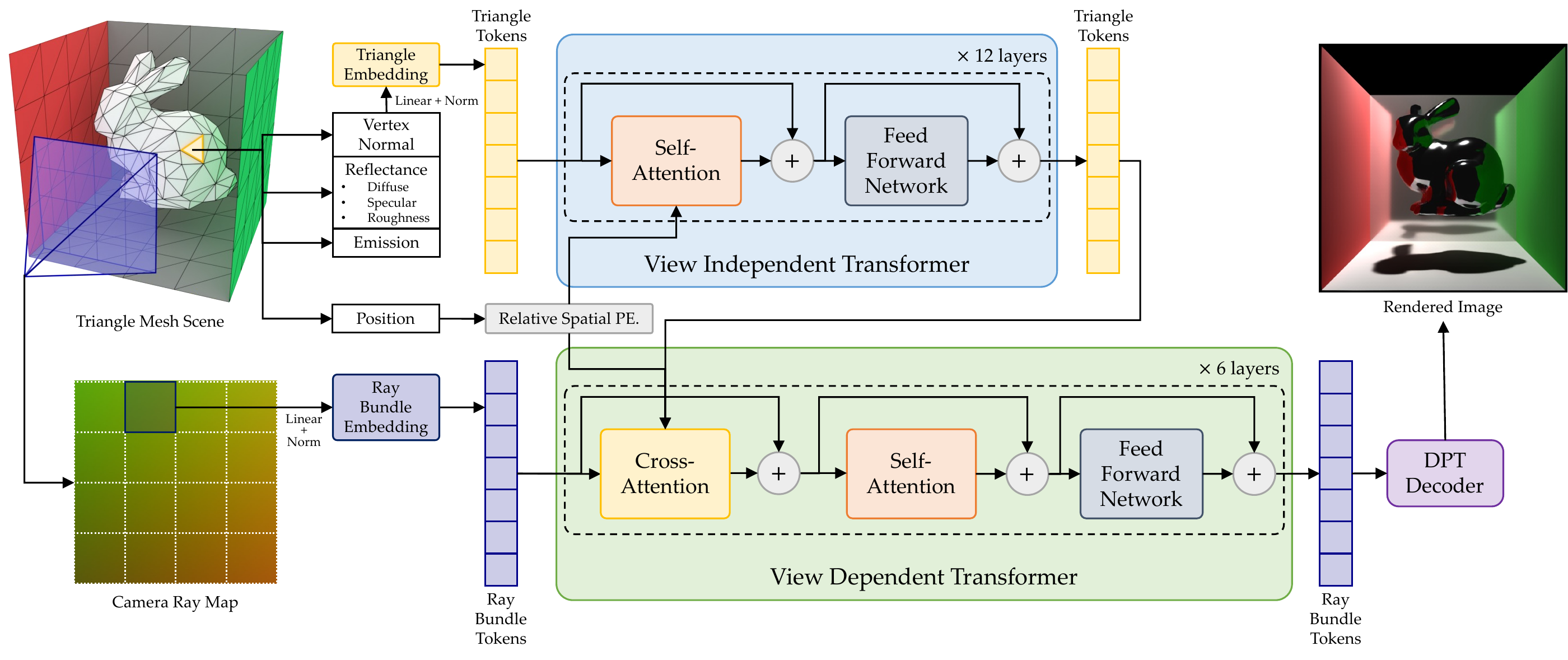}
    \caption{RenderFormer Architecture Overview. Top: the
      view-independent stage resolves triangle-to-triangle light
      transport from a sequence of triangle tokens that encode the
      reflectance properties of each triangle. The relative position
      of each triangle is separately encoded, and applied to each
      token at each self-attention layer.  Bottom: the view-dependent
      stage takes as input the virtual camera position encoded as a
      sequence of ray-bundles. Guided by the resulting triangle tokens
      from the view-independent stage via a cross-attention layer, the
      ray-bundle tokens are transformed to tokens encoding the
      outgoing radiance per view ray. Finally, the ray-bundle tokens
      are transformed to log-encoded HDR radiance value through an
      additional dense vision transformer.}
    \label{fig:arch}
\end{figure*}

RenderFormer is composed of two stages: a \emph{view-independent}
stage and a \emph{view-dependent} stage.  Both stages utilize the
transformer architecture.  The view-independent stage
(\autoref{sec:independent}) takes a sequence of triangles with
corresponding properties as input, and transforms it to a sequence of
per-triangle features that store a neural encoding of the triangle's
overall outgoing radiance.  The view-dependent stage
(\autoref{sec:dependent}) takes the transformed triangle sequence as
input as well as tokens that represent bundles of rays corresponding
to $8 \times 8$ pixel patches in the target image and transforms the
latter to outgoing radiance values corresponding to each ray in the
bundle.  We train RenderFormer end-to-end (\autoref{sec:training})
once, after which a triangle-based scene can be fed into RenderFormer
without any fine-tuning or training. 

\subsection{View-independent Stage}
\label{sec:independent}
\paragraph{Transformer Architecture}
The view-independent stage clo\-se\-ly follows the original
transformer architecture~\cite{Vaswani:2017:AIA} with full
bidirectional self-attention. The transformer takes as input a
sequence of triangle embeddings (\ie, tokens).  Each triangle token
encodes all relevant information for rendering such as surface normal
and reflectance. In addition we add $16$ register tokens to the input
sequence that can be used by the transformer to store global
information and potentially remove high-frequency noise in the
embedding~\cite{Darcet:2024:VNR}.  Each triangle and register token is
a $768$-dimensional vector.  The view-independent stage is composed of
$12$ transformer layers, where each layer has $6$ heads and $768$
hidden units, followed by a $768 \times 4$ feed-forward fully
connected network.  We follow LLaMA~\cite{Touvron:2023:LOE} and apply
pre-normalization using RMS-Normalization~\cite{Zhang:2019:RMS} and
use SwiGLU as activation function~\cite{Shazeer:2020:GVT}. Furthermore
we leverage QK-Normalization~\cite{Henry:2020:QKN} to stabilize
training. \autoref{fig:arch} (top) summarizes the architecture of the
view-independent stage.

\paragraph{Relative Spatial Positional Embedding}
A key difference between RenderFormer and typical uses of transformers
(\eg, large language models) is that the index position of the token
(\ie, triangle) in the sequence is irrelevant; swapping two triangles
in the sequence should produce the same result.  However, the position
of the triangle in the virtual world matters. The contribution to the
global light transport differs for two triangles with exactly the same
reflectance properties and shape (and thus with identical token
embedding) but at different positions in the scene.  Furthermore,
translating the whole scene (including light sources and virtual
camera) does not alter the light transport.  Hence, RenderFormer
requires a relative positional encoding based on the 3D spatial
location for each triangle with respect to other triangles.  We,
therefore, do not embed the \emph{absolute} position of the triangle
by adding the positional encoding directly to the triangle token, but
instead adapt Rotational Positional Encoding (RoPE)~\cite{Su:2024:RET}
to modify the triangle token to embed the triangle's \emph{relative}
3D spatial location. RoPE expresses the positional embedding as a
rotation and relies on the fact that the composition of two rotations
is equivalent to a relative rotation between both. However, unlike a
simple index in a sequence, the position of the triangle is determined
by three $3$D vertices of floating point values. We therefore first
concatenate all three vertex positions into a 9D vector and multiply
each element, duplicated $6$ times, with each of $6$ frequencies (with
scales exponentially distributed between $1$ and $5$:
$[1.0, 1.3797, 1.9037, 2.6265, 3.6239, 5.0]$), yielding a vector of
$54$ scaled frequencies. Following RoPE, we encode each coefficient as
the sine and cosine of the angle proportional to the scaled
frequencies, and create a block-diagonal rotation matrix where each
sine/cosine pair determines the rotation for each $2 \times 2$ block.
Each of the $6$ attention heads operates on $128$ coefficients of the
triangle token embedding ($6$ heads $\times 128 = 768)$.
Consequently, we apply the block-rotation only to the first $108$
coefficients for each head and leave the remaining $20$ coefficients
unchanged.  Similar to RoPE, we apply the relative spatial positional
embedding to the tokens at each attention layer.  Ideally we would
like the relative positional encoding to also be invariant to scene
rotations. However, because SO($3$) is not commutative, this is
difficult to achieve with RoPE.

To ensure that the register tokens are also invariant to scene
translations, we also apply relative spatial positional encoding on
the register tokens using the average position of all scene vertices.

\paragraph{Triangle Embedding}
For each triangle we want to embed all relevant information needed for
rendering, such as shading normals, reflectance properties, and
emission (in case of a light source).  As noted above, the position
and shape of the triangle will be encoded via relative spatial
positional embedding.

We store a normal per vertex that is interpolated (and normalized)
over the triangle using an absolute positional encoding of the
per-vertex normals. Practically, we encode all three normals with
(NeRF) positional encoding~\cite{Mildenhall:2021:NRS} with $6$
frequencies (using the same frequencies as for relative spatial
positional embedding), which are subsequently expanded to a
$768$-dimensional vector through a single linear layer followed by
RMS-Normalization.

We model surface reflectance with a microfacet BRDF model using a GGX
normal-facet distribution~\cite{Walter:2007:MMR} parameterized by
diffuse albedo, specular albedo, and roughness. We stack the
reflectance parameters as well as emission into a $10$ dimensional
vector ($3$D for all parameters except for roughness ($1$D)).  This
$10$-dimensional vector is expanded to a $768$-dimensional vector by a
single linear layer followed by RMS-Normalization.  The resulting
$768$-dimensional vector is added to the above normal embedding.

\subsection{View-dependent Stage}
\label{sec:dependent}

The goal of the view-dependent stage is to transform the triangle
tokens transformed by the previous view-independent stage to radiance
pixel values corresponding to a given virtual camera.  For performance
reasons, we encode an $8 \times 8$ radiance pixel patch in an output
token.  Inspired by Gao~\etal~\shortcite{Gao:2024:CCA} and
Jin~\etal~\shortcite{Jin:2024:LAL}, we specify the virtual camera to
the view-dependent transformer by encoding a bundle of $8 \times 8$
rays that pass through the centers of the pixels in the corresponding
output patch.

\paragraph{Transformer Architecture}
We follow a similar architecture as for the view-independent
transformer, except that it transforms a sequence of ray-bundle tokens
(instead of triangle tokens), and we only repeat the attention layers
$6$ times instead of $12$. Furthermore we precede each self-attention
layer with a cross-attention layer that connects the ray-bundle tokens
with the triangle tokens (including register tokens) from the
view-independent stage.  The role of the cross-attention layer is to
find the triangles related to the rays in the ray-bundle. As before,
each transformer layer has $6$ heads, $768$ hidden units, and a
$768 \times 4$ feed-forward network. We again employ SwiGLU
activations, QK-normalization, and RMS-Normalization.  Furthermore, we
found that the view-dependent stage requires higher precision (tf32)
than the view-independent stage (which uses bf16) to convergence
during training.  In addition, to decode the pixel-patch tokens into
$\log(x+1)$-encoded HDR RGB radiance values, we found that even though
the radiance observed for each view ray is independent of other view
rays, sharing information between view rays through self-attention
between ray-bundle tokens improves rendering accuracy
(\autoref{tab:ablation}, 3rd vs. 4th row).  Furthermore, we employ a
dense vision transformer~\cite{Ranftl:2021:VTD} on the features from
the last $4$ layers of the view-dependent transformer, to further
improve accuracy (\autoref{tab:ablation}, 1st vs. 2nd row) \emph{and}
reduce (but not fully eliminate) resolution dependence.
\autoref{fig:arch} (bottom) summarizes the view-dependent
architecture.

\begin{table}[t]
  \caption{Ablation study of different model variants and
    architectures. Due to computational constraints, all ablation
    studies are performed at $256 \times 256$ resolution. Layer
    configurations are denoted as: \texttt{\#view-independent +
      \#view-dependent} layers.}
  \label{tab:ablation}
  \footnotesize
  \begin{tabular}{l|cccc}
    \hline
    Variant & PSNR $\uparrow$ & SSIM $\uparrow$ & LPIPS $\downarrow$ & FLIP $\downarrow$ \\ \hline
    \textbf{full view-dependent stage} & \textbf{29.77} & \textbf{0.9526} & \textbf{0.05514} & \textbf{0.1751} \\
    w/o dpt & 29.75 & 0.9476 & 0.05519 & 0.1806 \\
    w/o self-attention & 29.70 & 0.9503 & 0.05703 & 0.1766 \\
    w/o dpt \& w/o self-attention & 29.15 & 0.9396 & 0.06407 & 0.1836 \\ \hline
    \textbf{camera space view-dep. stage} & \textbf{29.77} & \textbf{0.9526} & \textbf{0.05514} & \textbf{0.1751} \\
    world space view-dep. stage & 28.98 & 0.9420 & 0.06309 & 0.1904 \\ \hline
    \textbf{205M} / 768d tokens / 12 + 6 layers & \textbf{29.77} & \textbf{0.9526} & \textbf{0.05514} & \textbf{0.1751} \\
    143M / 768d tokens / 8 + 4 layers & 28.99 & 0.9444 & 0.06408 & 0.1873 \\
    71M / 512d tokens / 8 + 4 layers & 28.27 & 0.9356 & 0.07238 & 0.2032 \\
    45M / 384d tokens / 8 + 4 layers & 27.87 & 0.9295 & 0.07921 & 0.2075 \\ \hline
    12 + 6 layers  & 29.77 & 0.9526 & 0.05514 & 0.1751 \\
    9 + 9 layers & 30.11 & 0.9554 & 0.05121 & 0.1735 \\
    \textbf{6 + 12 layers} & \textbf{30.38} & \textbf{0.9560} & \textbf{0.05043} & \textbf{0.1685} \\
    0 + 18 layers  & 28.28 & 0.9355 & 0.07152 & 0.1994 \\ \hline
  \end{tabular}
\end{table}

To reduce the degrees of freedom in the training data, we perform the
view-dependent stage in camera coordinates, rather than in world
coordinates as in the view-dependent stage.  This is trivially
achieved by applying the relative positional spatial embedding using
transformed vertex coordinates at each attention layer.  We do not
apply any other transformation (\eg, normals) because after the
view-independent transformation, the interpretation of the triangle
tokens does not align anymore with the original embedding.  By
expressing the triangles' (and registers') positional embedding in
camera coordinates we avoid having to learn the world-to-camera
transformation, which also helps to improve accuracy
(\autoref{tab:ablation}, 5th-6th row).

\paragraph{Ray Bundle Embedding}
Each ray bundle is a collection of $8 \times 8$ rays that go through
the center of the pixels of the corresponding pixel patch.  Because
the scene is expressed in camera coordinates in the view-dependent
stage, the origin of all rays is $(0,0,0)$. We therefore, only need to
encode the normalized directions of each ray.  We stack the $64$
direction vectors in a $192$-dimensional vector which is subsequently
encoded by a single linear layer followed by RMS-Normalization into a
$768$-dimensional token.

\subsection{Training}
\label{sec:training}

We train RenderFormer end-to-end using the
AdamW~\cite{Loshchilov:2019:DWD} optimizer with a batch size of $128$,
a linear warm-up learning step size of $1.0 \times 10^{-4}$ for
$8,\!000$ steps followed by a cosine decay schedule on $8$ NVIDIA A100
GPUs with 40GB of memory using Flash-Attention 2~\cite{Dao:2024:FA2}
and Liger Kernel~\cite{Hsu:2024:LKE} for speed-up. We, first train
RenderFormer at $256 \times 256$ resolution with a maximum mesh size
of $1,\!536$ triangles for $500k$ iterations which took approximately
$5$ days, followed by $100k$ additional fine-tuning iterations at
$512 \times 512$ resolution with a maximum mesh size of $4,\!096$
triangles, which took an additional $3$ days.  While RenderFormer is
invariant to scene translations due to the relative spatial positional
embedding, it is not invariant to rotations. We therefore improve
stability to scene rotations by applying a random rotation to the
scene (including the camera); this does not require rerendering and
thus rotation is performed on-the-fly during training using
RoMa~\cite{Bregier:2021:DRM}.

\newcommand{\templateFigWidth}{0.1\textwidth}

\begin{figure}
\renewcommand{\arraystretch}{-0.1}
\addtolength{\tabcolsep}{-3.0pt}
 \begin{tabular}{cccc}
    \includegraphics[width=\templateFigWidth]{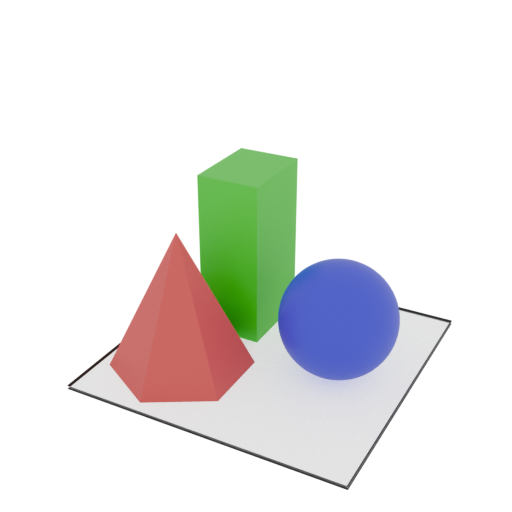}
    &\includegraphics[width=\templateFigWidth]{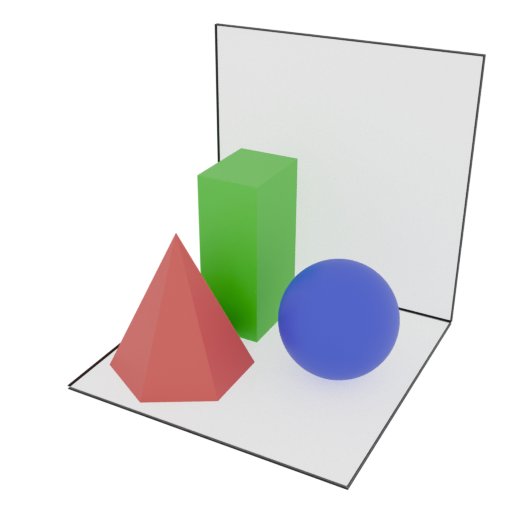} &
    \includegraphics[width=\templateFigWidth]{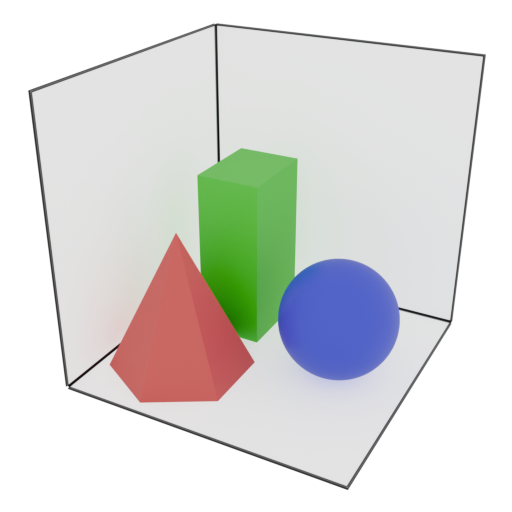}
    &\includegraphics[width=\templateFigWidth]{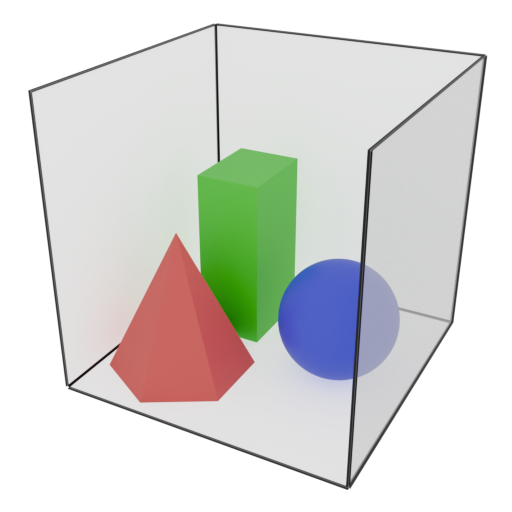} \\
 \end{tabular}
 \caption{The four template scenes used for generating training data.}
 \label{fig:templates}
\end{figure}

\paragraph{Loss Function}
We train RenderFormer in a supervised manner by computing the L1 loss
between a rendered reference HDR image of a synthetic scene and the
RenderFormer HDR prediction.  To avoid that small errors on bright
highlights dominate the loss, we first apply a log transform to the
images before computing the L1 loss.  In addition, to minimize
perceptual differences, we also include an LPIPS
loss~\cite{Zhang:2018:UED} on a tone-mapped version
($clamp(\log{I} / \log{2}, 0, 1)$) of both images.  The final loss is
then: $loss_{L1} + 0.05 loss_{LPIPS}$.

\paragraph{Training Data}
Our training set consists of synthetic scenes composed of $1$ to $3$
randomly selected objects from the Objaverse dataset
\cite{Deitke:2023:OAU} randomly placed in one of four template scenes
(shown in~\autoref{fig:templates}) that consist of a combination of
(randomly translated, rotated, and scaled) ground, back, and side
walls.  The camera is placed outside the scene such that its view is
not blocked by any of the template walls and with a field of view
(FOV) uniformly sampled between $30^\circ$ and $60^\circ$. The camera
is aimed towards the center (with some perturbations to avoid always
aiming at the exact center), at a distance uniformly sampled between
$1.5$ and $2.0$ units, where one unit corresponds to the size of the
scene's bounding box. Between $1$ to $8$ light sources (\ie, triangles
with a diffuse emission), with an intensity uniformly sampled between
$2,\!500$ and $5,\!000$ $W/units^2$, are placed following a similar
procedure as the camera, but with a distance uniformly sampled between
$2.1$ and $2.7$ units.  We randomly assign material parameters either
per-object or per-triangle with a 1:1 ratio.  We randomly assign an
RGB color to the diffuse albedo with maximum intensity per color
channel set such that the sum with the monochromatic specular albedo
lies between $0.9$ and $1.0$ (uniformly sampled).  Roughness is
log-sampled in $[0.01, 1.0]$.  Furthermore, we randomly select, with
equal probability, whether the object is shaded with per-vertex
normals or flat-shaded.

The runtime-complexity of attention layers scales quadratically with
the number of tokens, and thus triangles in our case.  As a result, we
limit the total number of triangles in our scenes to $4,\!096$;
increasing this limit is an interesting avenue for future research.
Since the objects in the Objaverse dataset easily exceed our triangle
budget, we remesh the objects by first removing interior or malformed
triangles (by converting to a signed distance field followed by a
marching cubes step to convert it back to a clean triangle mesh),
followed by Qslim to lower the number of faces between $256$ to
$3,\!072$.

We render $8$M HDR training images for $2$M synthetic scenes from $4$
different viewpoints at $256 \times 256$ resolution (with a maximum
triangle count of $1,\!536$), and an additional $8$M HDR training
images at $512 \times 512$ resolution with a maximum triangle count of
$4,\!096$ using Blender Cycles with $4,\!096$ samples per pixel (using
adaptive sampling and denoising).

\section{Results}
\label{sec:results}
\begin{figure*}
  {
    \footnotesize
    \newcommand{\resFigWidth}{0.133\textwidth}
    \newcommand{\resTextWidth}{0.11\textwidth}
    \renewcommand{\arraystretch}{0.5}
    \addtolength{\tabcolsep}{-5.5pt}
    \centering
    \begin{tabular}{ ccccp{0.2cm}cccc }
      & Reference & RenderFormer & Difference $5\times$ & & Reference & RenderFormer & Difference $5\times$ & \\

      \parbox{\resTextWidth}{
      \vspace{-2cm}
      \centering
        PSNR: 27.63\\
        SSIM: 0.9609\\
        LPSIPS: 0.04350\\
        FLIP: 0.08558
      }
      &\includegraphics[width=\resFigWidth]{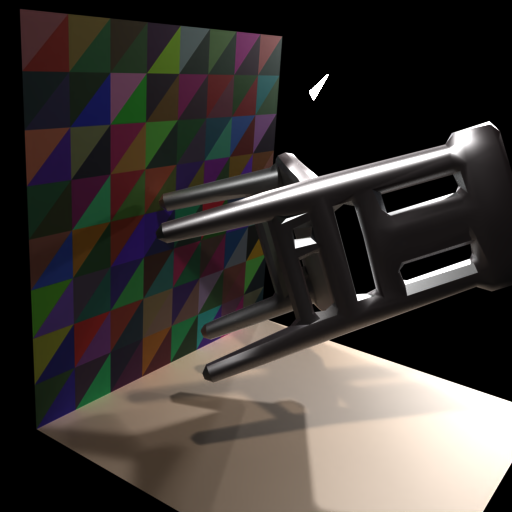}
      &\includegraphics[width=\resFigWidth]{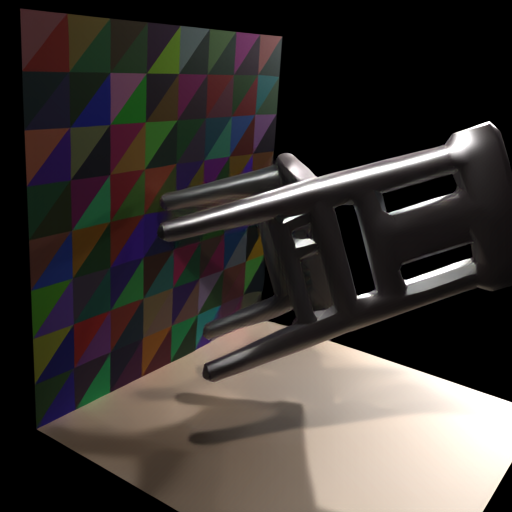}
      &\includegraphics[width=\resFigWidth]{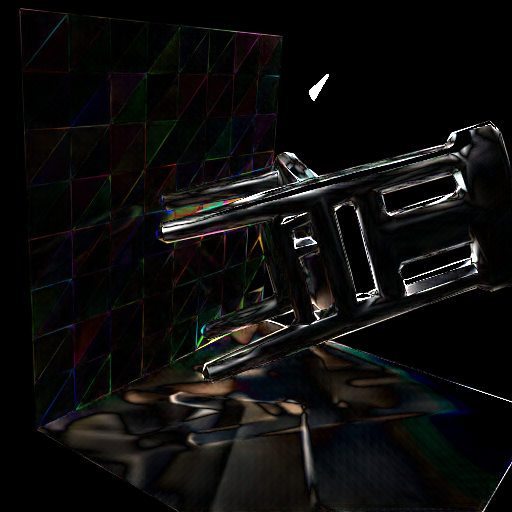}
      &
      &\includegraphics[width=\resFigWidth]{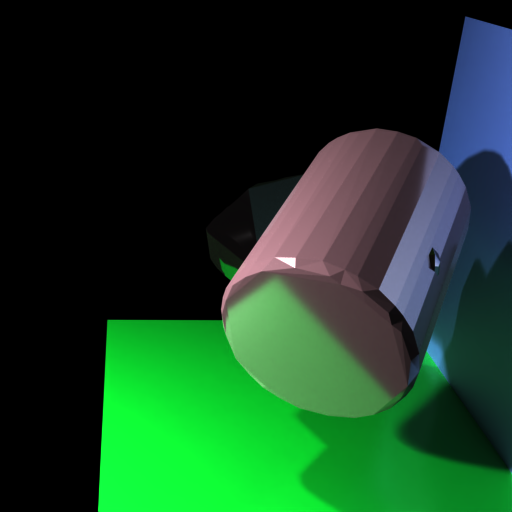}
      &\includegraphics[width=\resFigWidth]{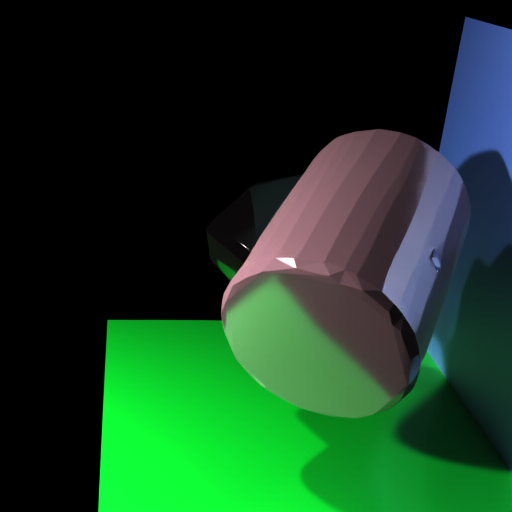}
      &\includegraphics[width=\resFigWidth]{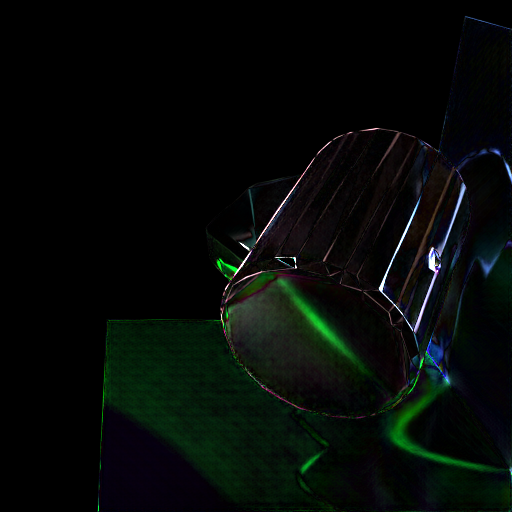}
      &
      \parbox{\resTextWidth}{
      \vspace{-2cm}
      \centering
        PSNR: 34.21\\
        SSIM: 0.9802\\
        LPIPS: 0.01361\\
        FLIP: 0.05515
      }\\

      \parbox{\resTextWidth}{
      \vspace{-2cm}
      \centering
        PSNR: 34.98\\
        SSIM: 0.9881\\
        LPIPS: 0.02247\\
        FLIP: 0.04377
      }
      &\includegraphics[width=\resFigWidth]{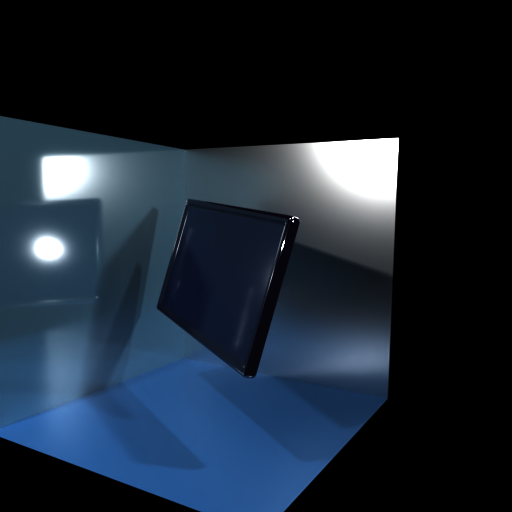}
      &\includegraphics[width=\resFigWidth]{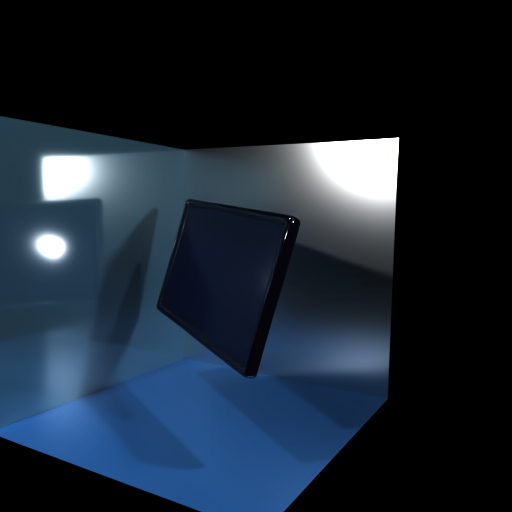}
      &\includegraphics[width=\resFigWidth]{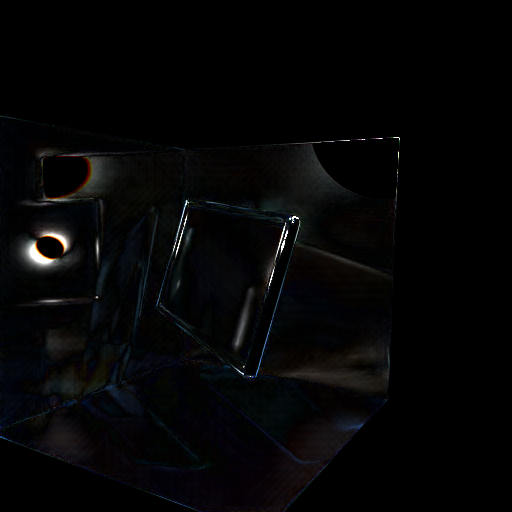}
      &
      &\includegraphics[width=\resFigWidth]{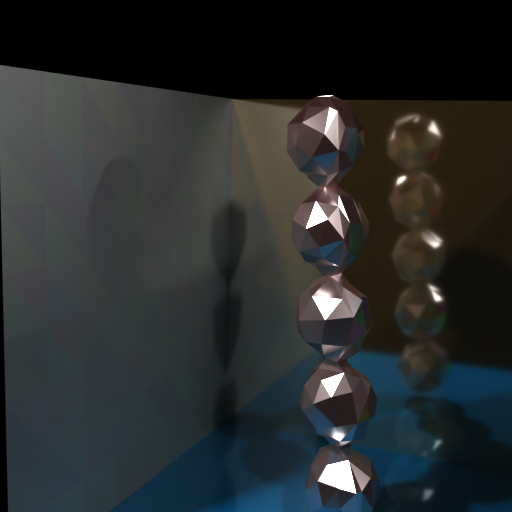}
      &\includegraphics[width=\resFigWidth]{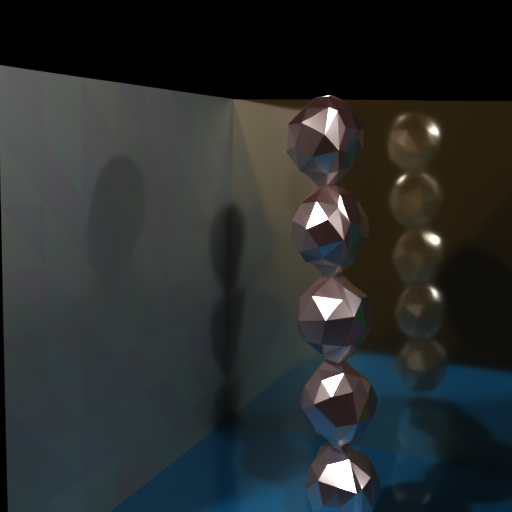}
      &\includegraphics[width=\resFigWidth]{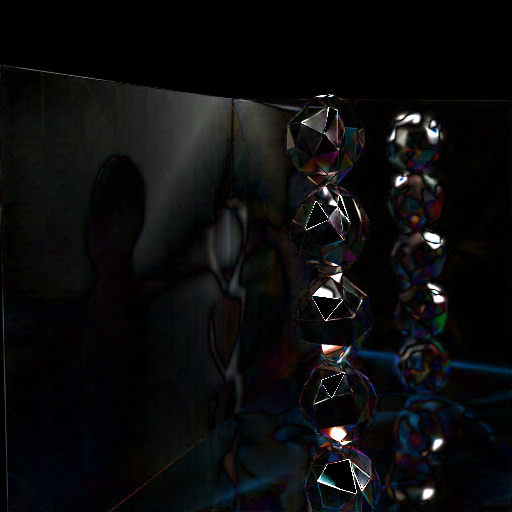}
      &
      \parbox{\resTextWidth}{
      \vspace{-2cm}
      \centering
        PSNR: 31.73\\
        SSIM: 0.9762\\
        LPSIP: 0.01537\\
        FLIP: 0.06922
      }
    \end{tabular}
  }
  \caption{A variety of scenes rendered with RenderFormer and compared
    to path-traced reference images.  We also list
    the PSNR, SSIM, LPIPS, and FLIP errors.}
  \label{fig:render}
\end{figure*}

\begin{table}[t]
  \caption{Timing comparison between RenderFormer and Blender Cycles
    with $4,\!096$ samples per pixels (matching the settings for
    training data generation).  We include both timings with and
    without adaptive sampling and denoising. In addition, we provide a
    breakdown of time spent in the view-independent and view-dependent
    stage.  Timings are measured in seconds with pre-cached kernels
    and excluding the cost of scene loading.}
  \label{tab:timing}
  \footnotesize
  \begin{tabular}{l|cccc}
    \hline
                             & \multicolumn{4}{c}{\autoref{fig:teaser}} \\
                             & First & Second & Third & Fourth \\
    \hline
    \#Triangles              & 5366  & 4400   & 4527  & 7321 \\
    \hline
    Cycles 4,096 adaptive spp + denoise & 3.97  & 4.73   & 3.77  & 2.71 \\
    Cycles 4,096 spp         & 12.05  & 11.21   & 9.95  & 7.83 \\ 
    RenderFormer             & 0.0760 & 0.0613 & 0.0625 & 0.0978 \\
    \hline
    View-independent stage   & 0.0282 & 0.0186 & 0.0192 & 0.0429 \\
    View-dependent stage     & 0.0478 & 0.0427 & 0.0433 & 0.0549 \\
    \hline
  \end{tabular}
\end{table}

We demonstrate RenderFormer on a variety of scenes
(\autoref{fig:teaser} and \autoref{fig:render}) showcasing different
aspects of global light transport.  For each example, we show a
reference render computed with Blender Cycles with $4,\!096$ samples
per pixel and a difference image scaled $5\times$ as well as the PSNR,
SSIM, LPIPS~\cite{Zhang:2018:UED} and
HDR-FLIP~\cite{Andersson:2020:FAD} errors.  Qualitatively, the
RenderFormer results look visually similar albeit not exactly the
same.  Nevertheless, RenderFormer manages to include many important
light transport effects such as shadows, diffuse and specular
interreflections, glossy reflections, and multiple specular
interreflections.  While not explicitly enforced, RenderFormer is
stable to changes in scene parameters as shown in the supplemental
video where we move the camera, move the lighting, and change
reflectance properties.

\begin{figure}
  \newcommand{\timingExtFigWidth}{0.121\textwidth}
  \renewcommand{\arraystretch}{0.0}
  \addtolength{\tabcolsep}{-5.5pt}
  {
  \footnotesize
  \begin{tabular}{ cccc }
    \includegraphics[width=\timingExtFigWidth]{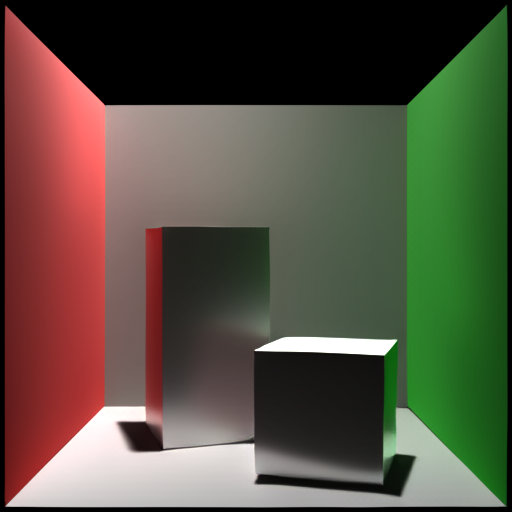} & 
    \includegraphics[width=\timingExtFigWidth]{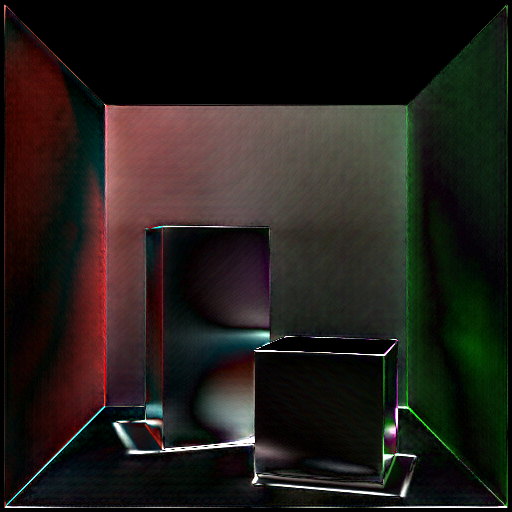} & 
    \includegraphics[width=\timingExtFigWidth]{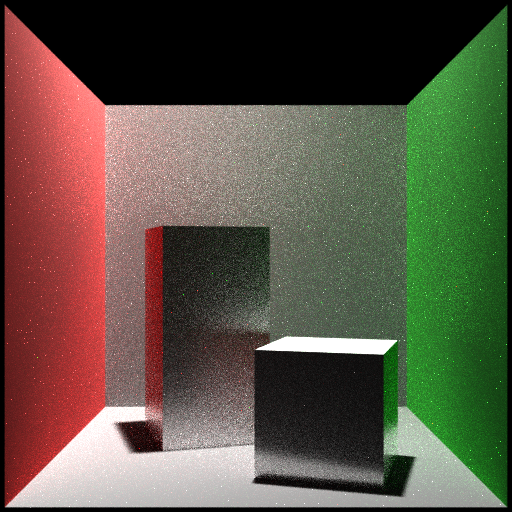} & 
    \includegraphics[width=\timingExtFigWidth]{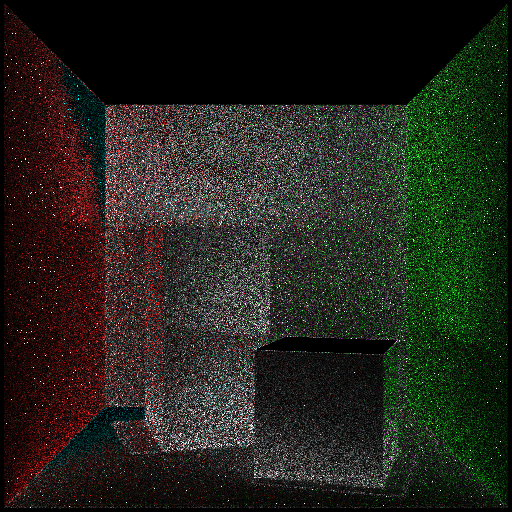} \\
    \vspace{0.1cm}\\
    RenderFormer & Error $\times5$ & Path Tracing 26 spp & Error $\times5$\\
  \end{tabular}
  }
  \caption{Equal-time comparison between RenderFormer and Blender
    Cycles with (non-adaptive) $26$ sampler-per-pixel and without denoising.}
  \label{fig:equaltime}
\end{figure}

\autoref{tab:timing} compares the timings on the four scenes
in~\autoref{fig:teaser} of our unoptimized RenderFormer (pure PyTorch
implementation without DNN compilation, but with pre-caching of
kernels) and Blender Cycles with $4,\!096$ samples per pixel (matching
RenderFormer's training data) at $512 \times 512$ resolution on a
single NVIDIA A100 GPU.  To provide further insight, we also provide a
qualitative equal-time comparison (\autoref{fig:equaltime}) of the
first scene from~\autoref{fig:teaser}; because the fixed cost of
denoising exceeds the RenderFormer times and the scene-dependent
non-linear cost of adaptive sampling, we disable both optimizations
for the equal-time comparison.  Besides optimizing the RenderFormer
implementation, we can further speed up rendering for static scenes by
reusing the view-independent transformed sequence or for animations by
rendering $48$ frames in parallel by batching.

\begin{figure*}
  \newcommand{\ablationFigWidth}{0.202\textwidth}
  \renewcommand{\arraystretch}{0.0}
  \addtolength{\tabcolsep}{-5.5pt}
  {
  \footnotesize
  \begin{tabular}{ ccccc }
    \includegraphics[width=\ablationFigWidth]{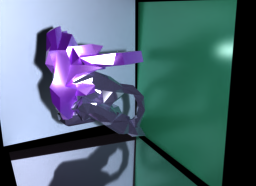}
    &\includegraphics[width=\ablationFigWidth]{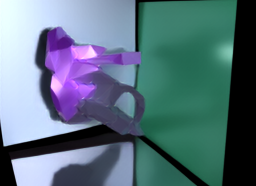}
    &\includegraphics[width=\ablationFigWidth]{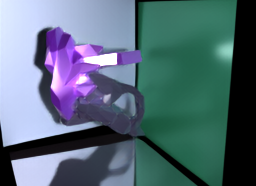}
    &\includegraphics[width=\ablationFigWidth]{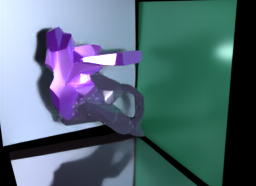}
    &\includegraphics[width=\ablationFigWidth]{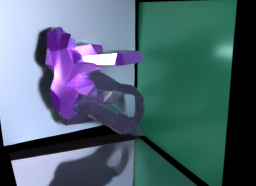} \\
    \vspace{-0.2cm} \\
    \textcolor{white}{Reference} & \textcolor{white}{0+18} & \textcolor{white}{6+12}& \textcolor{white}{9+9} & \textcolor{white}{\bf{12+6} }
  \end{tabular}
  }
  \caption{Qualitative comparison of varying
    \texttt{\#view-independent + \#view-dependent} attention layers per
    stage.  RenderFormer is shown in the last column with a ratio of
    $12$ view-independent versus $6$ view-dependent layers.}
  \label{fig:ablation}
\end{figure*}

\subsection{Analysis \& Ablation Study}
\label{sec:analysis}
RenderFormer's architecture differs from prior neural rendering
methods, and it follows a significantly different way of solving the
rendering equation compared to classic global illumination methods.
To gain more insight on the inner-workings of RenderFormer we perform
a series of ablation studies.  Due to computational constraints, we
perform all ablation experiments at $256 \times 256$ resolution.

\paragraph{Relative Spatial Positional Embedding}
One of the main differences between RenderFormer and traditional
transformers is the positional encoding based on the position of the
triangles in world space instead of the sequence index position, using
a novel relative spatial positional encoding based on RoPE. However,
we also used a NeRF-like positional encoding for embedding the vertex
normals.  This raises the questions whether it would be possible to
also embed the triangle positions together with the normals using a
NeRF positional encoding instead.  However, we found that training
with such positional encoding for the triangles' positions is not
stable, and it is prone to converge to a suboptimal local minimum.

\paragraph{Model Size}
A key design parameter in transformer models is the token feature
length; larger features result in larger models, and thus longer
training time. \autoref{tab:ablation} ($7$th to $11$th row) lists
average PSNR, SSIM, LPIPS~\cite{Zhang:2018:UED} and
HDR-FLIP~\cite{Andersson:2020:FAD} errors over $400$ test scenes
rendered from $4$ viewpoints (\ie, $1,\!600$ total) for models trained
with feature lengths ranging from $768$ to $192$. We also adjust the
number of attention layers to further reduce the number of model
parameters from $205$M to $143$M, $71$M, and $45$M parameters
respectively.  In general, more parameters yield more accurate
results.

\paragraph{Number of Layers}
In the previous experiment, we purposefully kept the ratio of
attention layers between the view-independent and view-dependent stage
constant.  We perform an additional ablation experiment to better
understand the impact of the ratio of attention layers between both
stages.  \autoref{tab:ablation} (rows $11$-$14$) compares RenderFormer
models with a different subdivisions of a total of $18$ attention
layers over the two stages. \autoref{fig:ablation} qualitative shows
the impact of varying the attention layers per stage.  We observe that
RenderFormer benefits from including more attention-layers in the
view-dependent stage than in the view-independent stage.  Fully
eliminating the view-independent stage (\autoref{tab:ablation}, $14$th
row and \autoref{fig:ablation}, $2$nd column) fails to produce good
results, indicating that the view-independent stage is necessary for
obtaining good results.  However, rendering often requires a careful
balance between accuracy and speed.  The runtime of each stage depends
on different factors.  The view-independent stage scales roughly by
$\mathcal{O}(\#tris^2)$, whereas the view-dependent layers scales by
$\mathcal{O}(\#bundles^2 + \#bundles \times \#tris)$.  Furthermore,
the difference in precision (bf16 versus tf32) imposes an additional
hardware-dependent performance scale between both stages. The ideal
number of attention layers per stage is complex and depends on mesh
size, resolution, and hardware.  We therefore opt for a $12+6$ split
between view-independent and view-dependent attention layers, balancing
accuracy and training/render speed (\ie, $\sim\!25\%$ faster for
$\sim\!5\%$ loss in accuracy).

\newcommand{\viewIndepFigWidth}{0.12\textwidth}
\begin{figure}
\renewcommand{\arraystretch}{0.0}
\addtolength{\tabcolsep}{-5.5pt}
\begin{tabular}{ ccccc }
 \includegraphics[width=\viewIndepFigWidth]{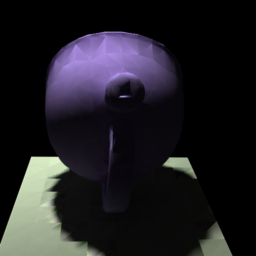}
 &\includegraphics[width=\viewIndepFigWidth]{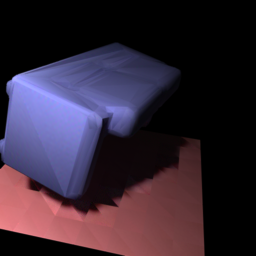}
 &\includegraphics[width=\viewIndepFigWidth]{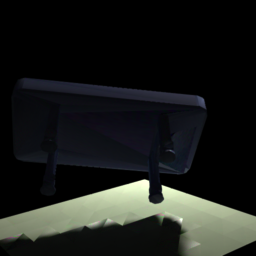}
  &\includegraphics[width=\viewIndepFigWidth]{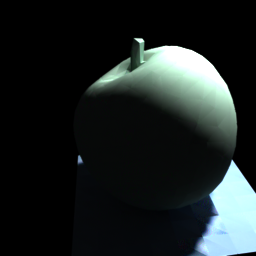} \\
  \end{tabular}
  \caption{Visualization of the transformed tokens from the
    view-independent stage that (after transformation) encode smooth
    diffuse shading and interreflections, as well as shadows at
    sub-triangle granularity.}
  \label{fig:viewindependent}
\end{figure}

\begin{figure}
  {
    \footnotesize
    \newcommand{\viewDepFigWidth}{0.16\textwidth}
    \renewcommand{\arraystretch}{0.0}
    \addtolength{\tabcolsep}{-5.5pt}
    \begin{tabular}{ ccc }
      \includegraphics[width=\viewDepFigWidth]{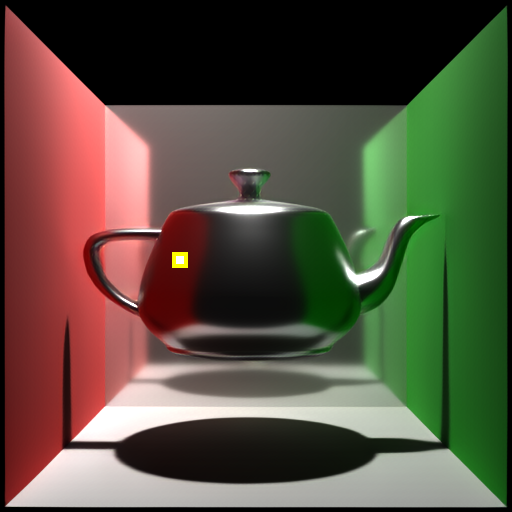} &
      \includegraphics[width=\viewDepFigWidth]{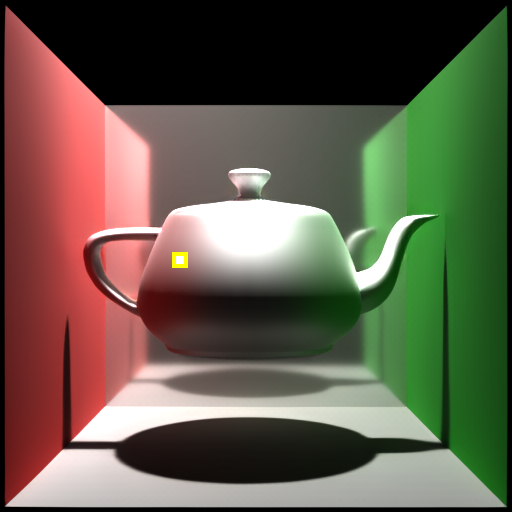} &
      \includegraphics[width=\viewDepFigWidth]{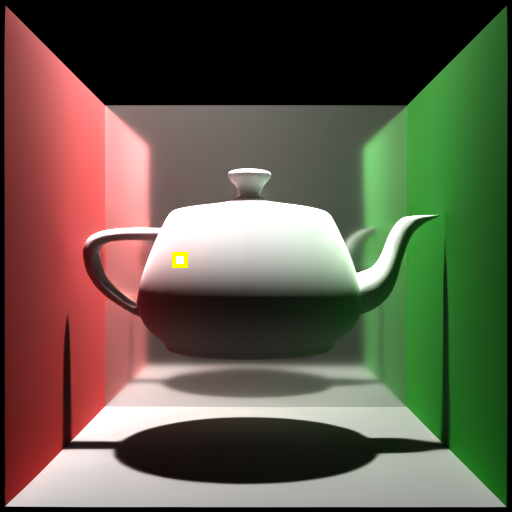} \\
      \includegraphics[width=\viewDepFigWidth]{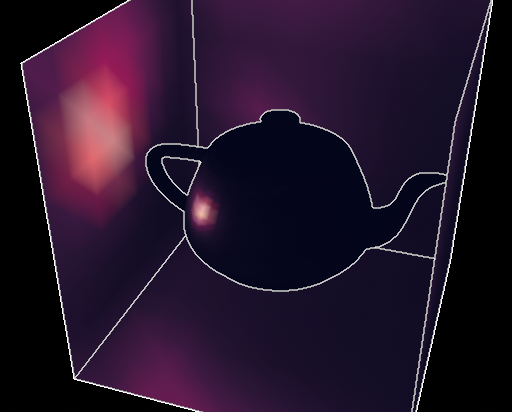} &
      \includegraphics[width=\viewDepFigWidth]{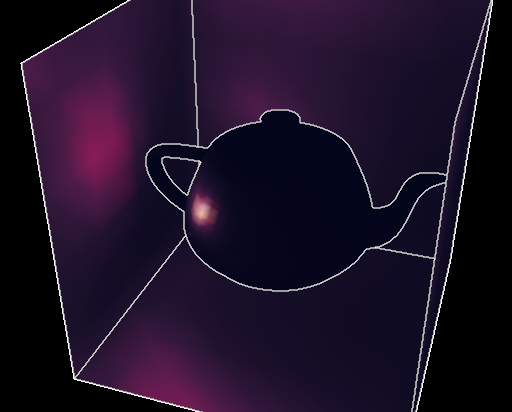} &
      \includegraphics[width=\viewDepFigWidth]{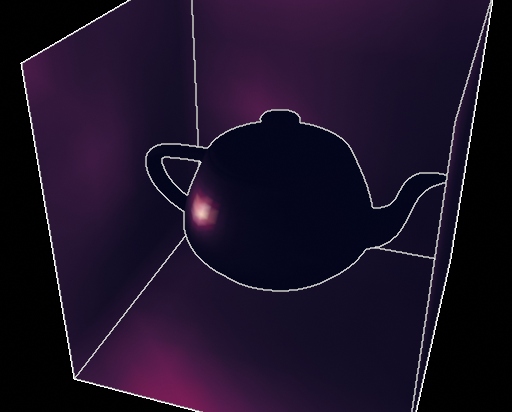} \\
      \vspace{0.1cm}\\
      roughness=0.3 & roughness=0.7 & roughness=0.99 \\
    \end{tabular}
  }
  \caption{Visualization of the average attention per triangle for
    a given ray-bundle in the view-dependent stage.  The average
    attention gives an indication on which triangles RenderFormer uses
    for computing the outgoing radiance for the rays in the bundle. As
    expected directly visible triangles and triangles around the
    reflected direction receive the most attention.}
  \label{fig:viewdependent}
\end{figure}

\paragraph{Role of the Different Stages}
The previous ablation study clearly shows that both stages are
necessary and each serve a role in the rendering pipeline.  However,
the previous ablation experiments do not give insights on what exactly
each stage does.

The interpretation of the embedding of the triangle tokens does not
follow the initial embedding after passing through the
view-independent stage, precluding direct visualization of the
triangle tokens. We therefore train an small auxiliary MLP that casts
a transformed triangle token into a $32 \times 32$ RGB texture for
each triangle.  Because the register tokens might include important
information, we also include a cross-attention layer between each
triangle token and the $16$ register tokens.  We train the MLP and the
cross-attention layers, while keeping the view-independent stage
frozen, on a small batch ($\sim500$) of simple solid colored diffuse
scenes.  The MLP is intentionally kept shallow to limit it to simple
operations that directly visualize the information embedded in the
tokens.  \autoref{fig:viewindependent} shows that the view-independent
stage resolves a significant portion of diffuse light transport
between triangles as well as shadows.

The view-dependent stage gathers information from the triangle tokens
to compute the observed radiance for each pixel.  In
\autoref{fig:viewdependent}, we visualize the (sum of the) attention
weights projected on their respective triangles for selected
ray-bundles and visualized from an appropriate view.  This
visualization shows how much each triangle contributes to the final
radiance observed for the given ray-bundle.  From
\autoref{fig:viewdependent} we can see that the main weight lies on
the directly visible triangle, as well as triangles around the
reflected direction.  We can also see that the weight distribution
changes when we increase the roughness of the material.

\newcommand{\triNumFigWidth}{0.12\textwidth}
\begin{figure}
  {
    \footnotesize
    \renewcommand{\arraystretch}{0.0}
    \addtolength{\tabcolsep}{-5.5pt}
    \begin{tabular}{ ccccc }
      \includegraphics[width=\triNumFigWidth]{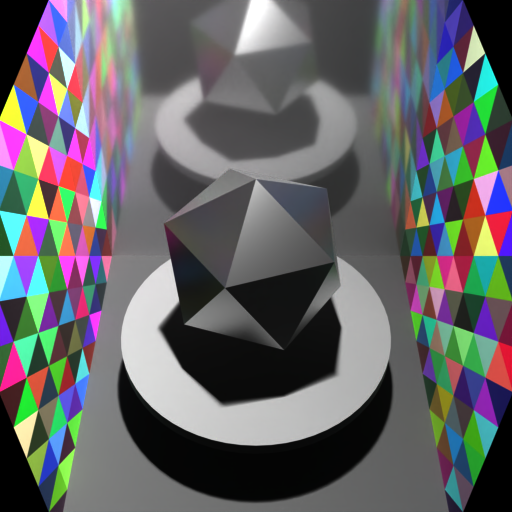}
      &\includegraphics[width=\triNumFigWidth]{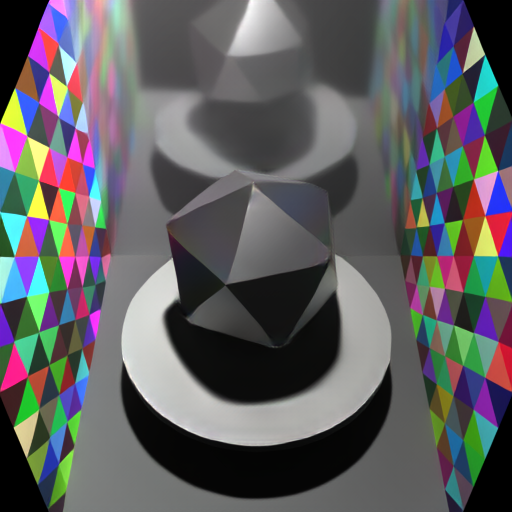}
      &\includegraphics[width=\triNumFigWidth]{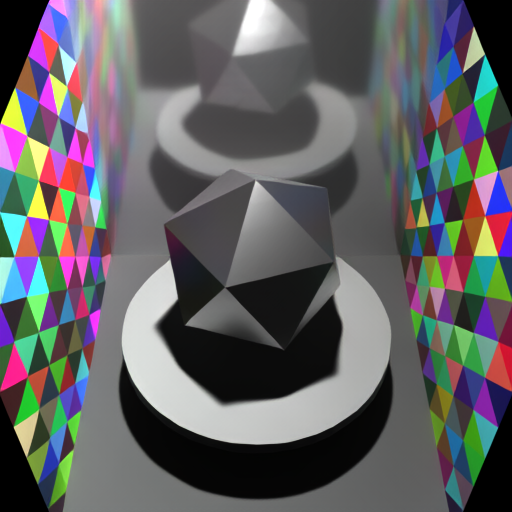}
      &\includegraphics[width=\triNumFigWidth]{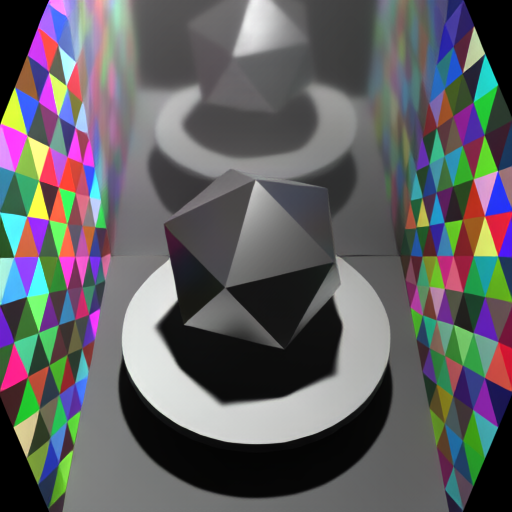} \\
      \vspace{0.1cm}\\
      Reference & Large Size. & Average Size & Average Size \\
                & $785$ Tris. & $1,\!831$ Tris. & $6,\!603$ Tris. \\
    \end{tabular}
  }
  \caption{Using larger than normal triangles for the pedestal and
    icosphere results in degraded shadows and shading ($2$nd
    column). Interestingly, this degradation is also visible in the
    reflections in the back wall.
  }
  \label{fig:triangles}
\end{figure}

\begin{figure*}
  {
    \footnotesize
    \newcommand{\numLightFigWidth}{0.1428\textwidth}
    \renewcommand{\arraystretch}{0.0}
    \addtolength{\tabcolsep}{-5.5pt}
    \begin{tabular}{ ccccccc }
      \includegraphics[width=\numLightFigWidth]{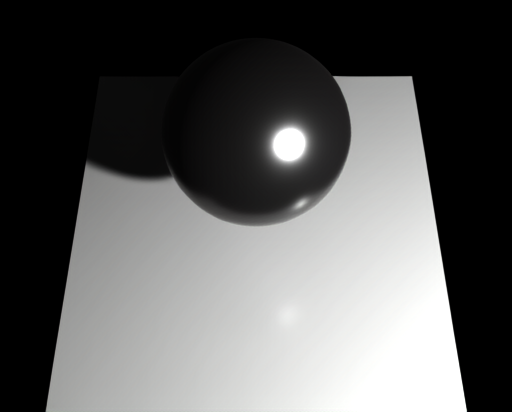}
      &\includegraphics[width=\numLightFigWidth]{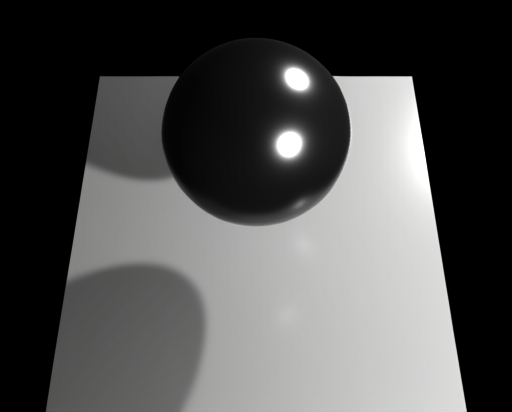}
      &\includegraphics[width=\numLightFigWidth]{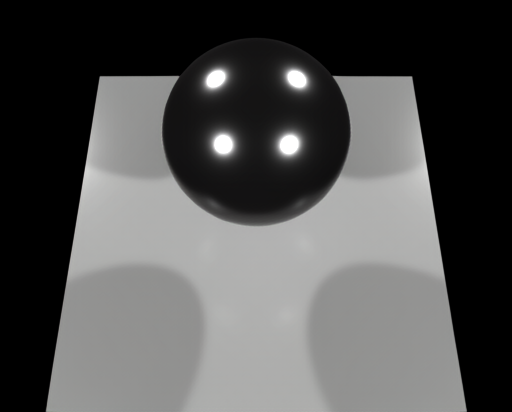}
      &\includegraphics[width=\numLightFigWidth]{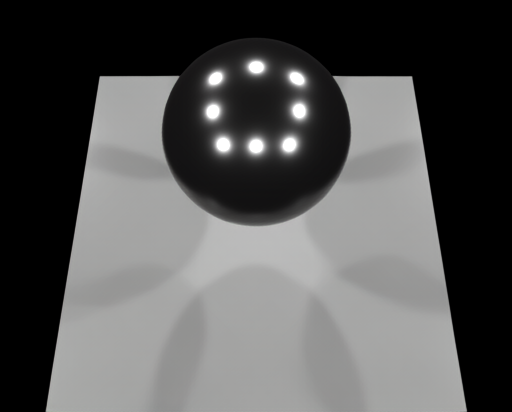}
      &\includegraphics[width=\numLightFigWidth]{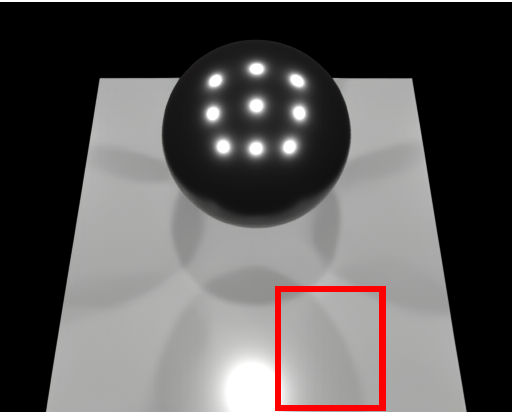}
      &\includegraphics[width=\numLightFigWidth]{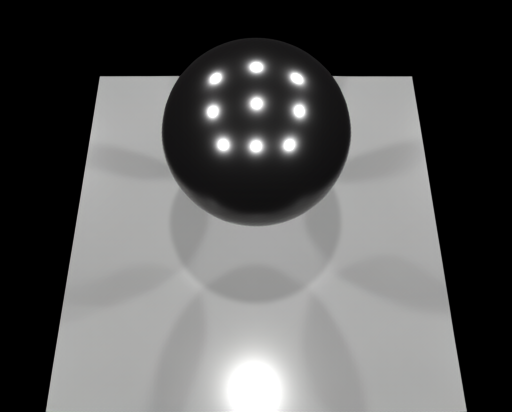}
      &\includegraphics[width=\numLightFigWidth]{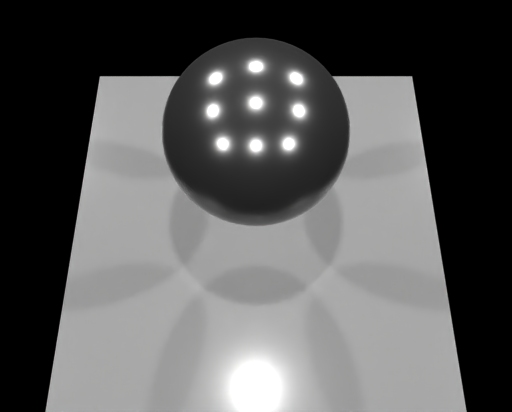} \\
      \vspace{0.1cm}\\
      1 Light & 2 Lights & 4 Lights & 8 Lights & 9 Lights (Direct) & 9 Lights (Sum) & 9 Lights (Reference) \\
    \end{tabular}
  }
  \caption{RenderFormer can handle multiple light sources with correct
    reflections and shadows ($1$st to $4$th column) as long as the
    number of lights does not exceed $8$ (as seen in training). For
    more lights, highlights or shadow might be missing (\eg, the
    missing double shadow at the bottom shadow in the $5$th
    column). In such a case, we can still compute a correct result by
    compositing multiple single-light images ($6$th column).}
  \label{fig:numlights}
\end{figure*}

\begin{figure*}
  \newcommand{\lightingLimitFigWidth}{0.125\textwidth}
  \renewcommand{\arraystretch}{0.0}
  \addtolength{\tabcolsep}{-5.5pt}
  {
  \footnotesize
  \begin{tabular}{cc|cc|cccc}
  \includegraphics[width=\lightingLimitFigWidth]{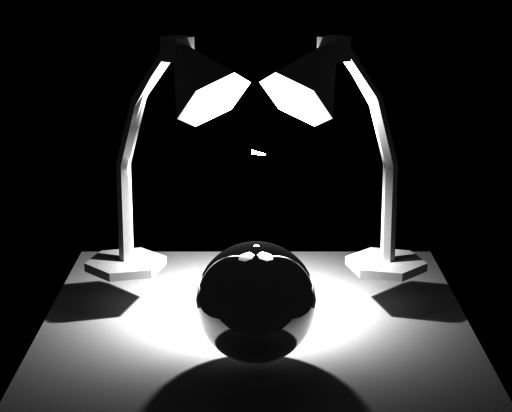} &
  \includegraphics[width=\lightingLimitFigWidth]{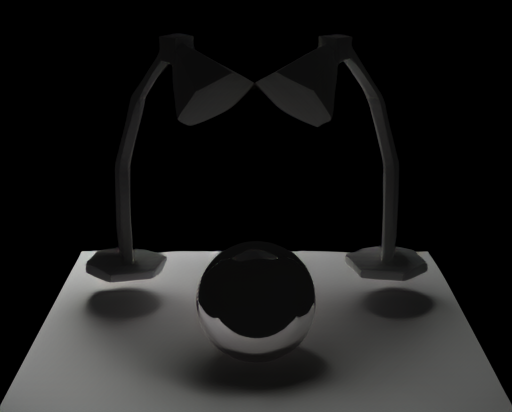} &
  \includegraphics[width=\lightingLimitFigWidth]{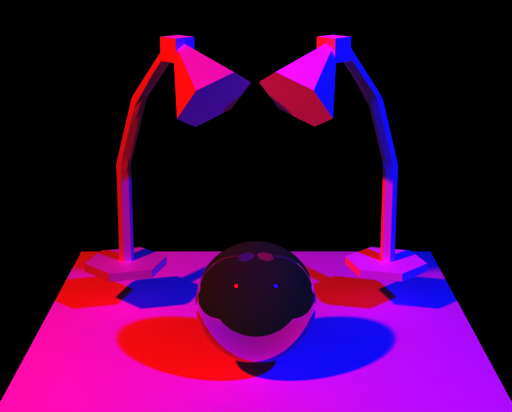} &
  \includegraphics[width=\lightingLimitFigWidth]{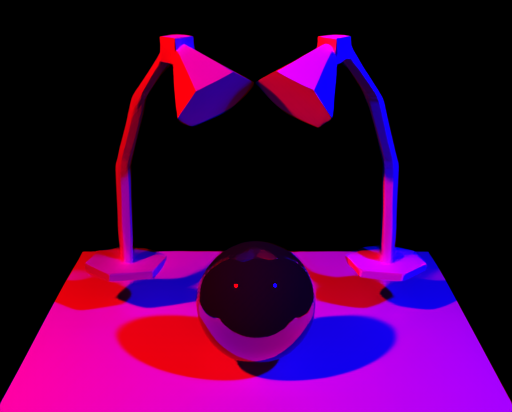} &
  \includegraphics[width=\lightingLimitFigWidth]{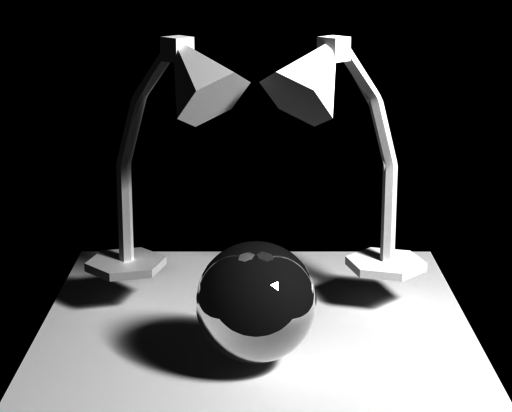} &
  \includegraphics[width=\lightingLimitFigWidth]{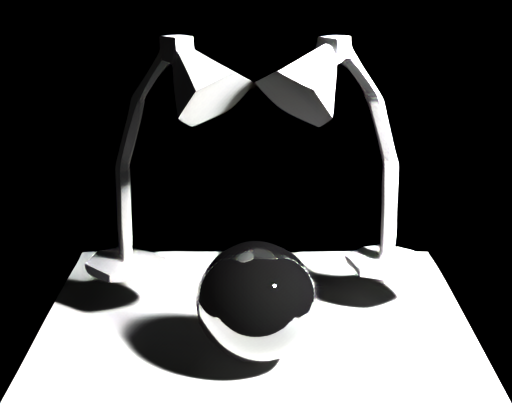} &
  \includegraphics[width=\lightingLimitFigWidth]{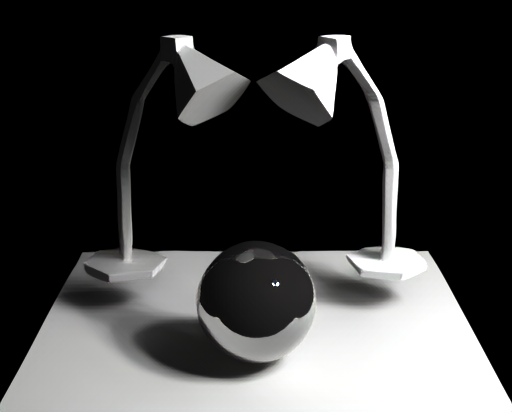} &
  \includegraphics[width=\lightingLimitFigWidth]{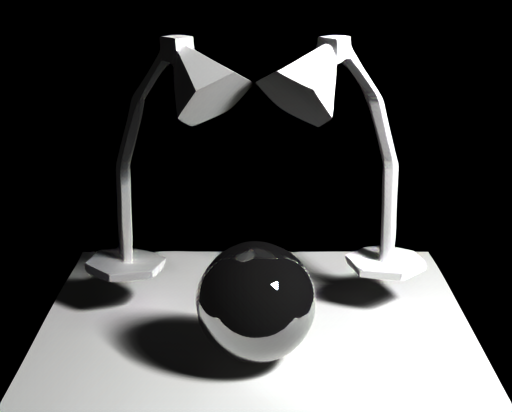} \\
    \vspace{0.1cm} \\
    Reference & RenderFormer & Reference & Composite & Reference & Single Triangle & Subdivided (16) & Composite (16) \\
    \multicolumn{2}{c|}{Light Inside Scene} & \multicolumn{2}{|c|}{Colored Light} & \multicolumn{4}{|c}{Large Light} 
  \end{tabular}
  }
  \caption{Left: RenderFormer was never trained with lights inside the
    scene, and thus fails to correctly render such scenes.  Middle:
    RenderFormer can simulate colored lights by leveraging linearity
    of light transport and blending three images (one for each color
    channel). Right: RenderFormer fails to correctly render scenes
    with light sources larger than those encountered during
    training. Subdividing the triangle can correct the error if the
    number of light sources does not exceed the maximum seen during
    training (8), in which case we can still leverage linearity of
    light transport by rendering each subdivided light separately.}
  \label{fig:lights}
\end{figure*}

\newcommand{\sizeNumFigWidth}{0.1175\textwidth}
\begin{figure}
  {
    \footnotesize
    \renewcommand{\arraystretch}{0.0}
    \addtolength{\tabcolsep}{-5.5pt}
    \begin{tabular}{ p{3mm}ccccc }
      \rotatebox{90}{\quad\quad\ \ Reference}
      &\includegraphics[width=\sizeNumFigWidth]{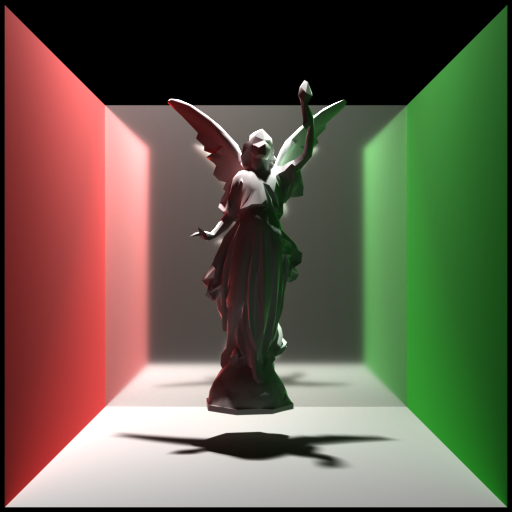}
      &\includegraphics[width=\sizeNumFigWidth]{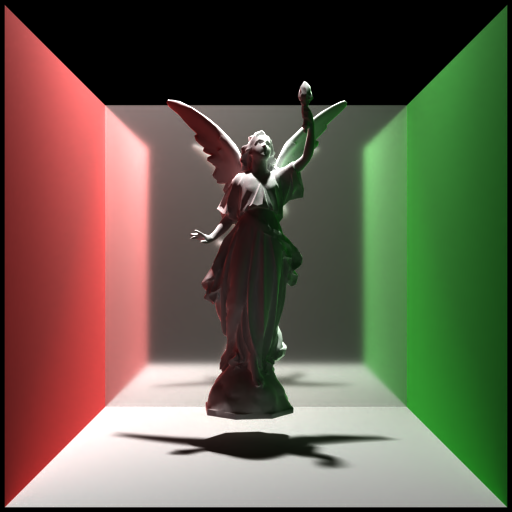}
      &\includegraphics[width=\sizeNumFigWidth]{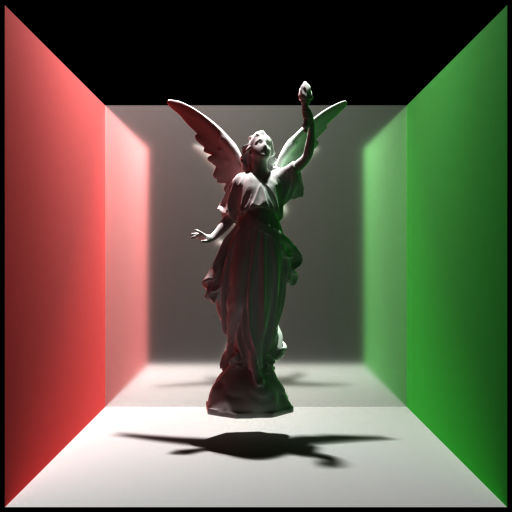}
      &\includegraphics[width=\sizeNumFigWidth]{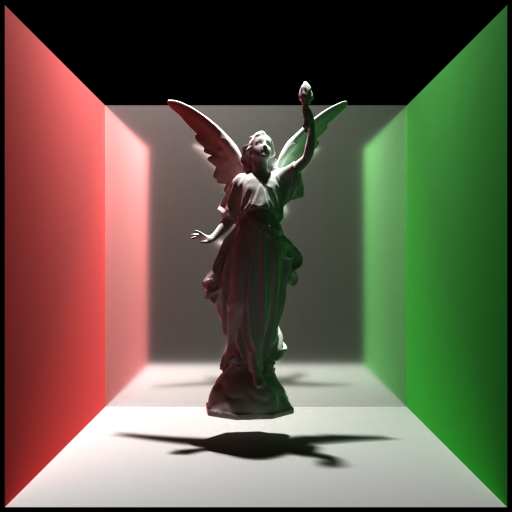} \\
      \rotatebox{90}{\quad RenderFormer}
      &\includegraphics[width=\sizeNumFigWidth]{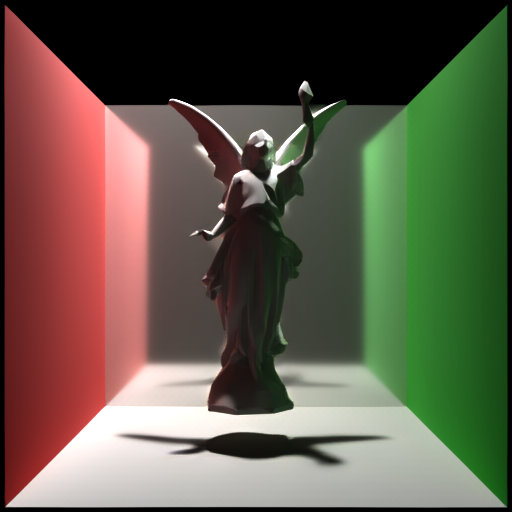}      
      &\includegraphics[width=\sizeNumFigWidth]{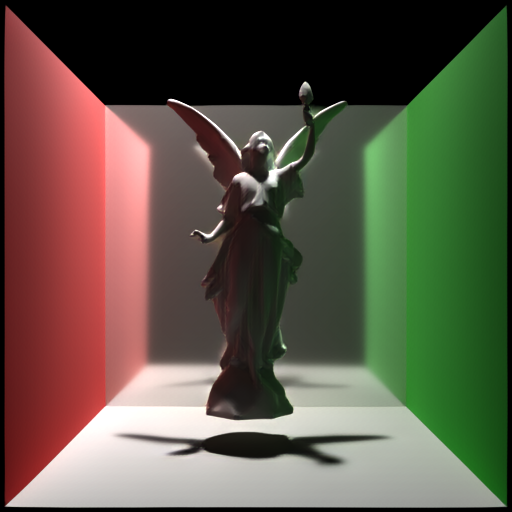}
      &\includegraphics[width=\sizeNumFigWidth]{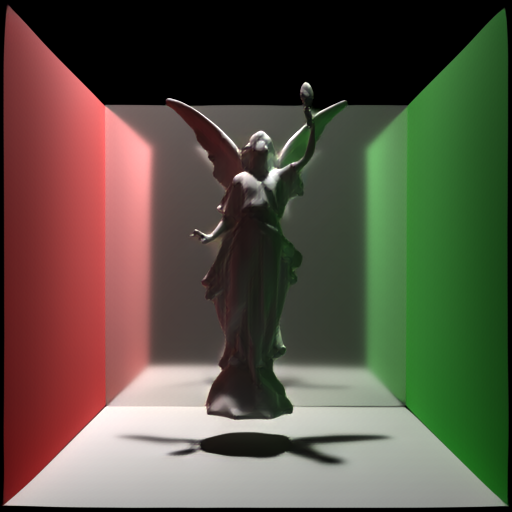}
      &\includegraphics[width=\sizeNumFigWidth]{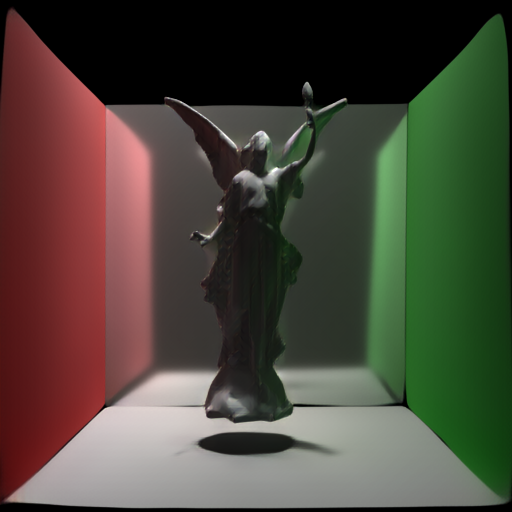} \\
      \vspace{0.1cm}\\
      & 3k & 11k & 23k & 45k \\
    \end{tabular}
  }
  \caption{RenderFormer can handle scenes with more triangles than for
    which it was trained, albeit with loss of detail and thin
    features.}
  \label{fig:size}
\end{figure}

\subsection{Generalization of Scene Parameters}
\label{sec:discussion}
While the previous ablation study and analysis provides more insight
on the inner-workings of RenderFormer, it does not give an indication
on how the model performs in practical situations and what its limits
are. Therefore, we perform several experiments to probe RenderFormer's
generalization capabilities with respect to the triangle mesh, light
sources, and camera.

\paragraph{Triangle Mesh}
Currently the training data is generated such that the triangles are
all roughly the same size. To better understand if the triangle size
affects the accuracy of the solution, we perform an experiment where we
render a scene twice (\autoref{fig:triangles}), once where the
pedestal and icosphere are represented by $1,\!318$ triangles of
average size, and once with $332$ larger triangles. As can be seen in
\autoref{fig:triangles} ($2$nd column), the quality of the shading and
shadows over larger triangles degrades due to the fact that the
triangle-tokens now need to store more complex information per
triangle.

The attention layers that form the core of transformers are costly in
terms of compute resources. Therefore, the training set for
RenderFormer is limited to scenes with at most $4,\!096$ triangles.
\autoref{fig:size} shows that RenderFormer can handle larger
triangle-meshes at inference time, albeit with some loss of
details. However, we observe that overall RenderFormer fails
gracefully with most of the light transport correctly modeled.  We
exploit this property during training by first pretraining on smaller
scenes ($1,\!536$ triangles) and then refine on larger scenes
($4,\!096$ triangles).  However, due to the $\mathcal{O}(N^2)$
complexity of the attention layers, there is a limit to how far the
model can be pushed before running out of
resources.  

Many attention-optimization techniques used in LLMs and Vision
Transformers~\cite{Dong:2023:ASD,Liu:2021:STH,Wang:2021:PVT} leverage
properties inherent to 1D sequences or 2D images whereas RenderFormer
operates in 3D, making direct application difficult. Adapting
state-of-the-art techniques from LLMs and Vision Transformers (such as
linear attention mechanisms, native sparse attention, and sequence
parallelism) is a promising avenue for future research, especially
when combined with established computer graphics methodologies such as
LoD and BVH.

\paragraph{Lighting}
RenderFormer is currently trained for scenes with $1$ to $8$ light
sources.  \autoref{fig:numlights} (columns $1$ to $4$) shows a
sequence of images of a scene with an increasing number of light
sources.  We observe that RenderFormer does not reliably handle cases
where the number of light sources exceeds the maximum seen during
training causing incomplete shadows or missing highlights
(\autoref{fig:numlights}, $5$th column).  The maximum number of light
sources can either be increased by training with more lights, or by
exploiting linearity of light transport by rendering each light source
separately and adding the resulting images (\autoref{fig:numlights},
$6$th column).

Currently, RenderFormer is also trained with light sources positioned
outside the scene, and placing a light source in the scene yields an
incorrect result (\autoref{fig:lights}, $2$nd column). Moreover,
RenderFormer is also trained for white light sources only;
RenderFormer ignores the color when it encounters a colored
light. This problem can be solved by either expanding the training set
or by leveraging linearity of light transport (\autoref{fig:lights},
$4$th column).

In addition, RenderFormer is trained for a limited range in light
source size. As expected, exceeding the trained light source size
(\autoref{fig:lights}, $6$th column) does not produce the correct
shadows. We can either expand the training set, or construct larger
light sources by subdividing the light source in more (smaller)
triangles (\autoref{fig:lights}, $7$th (direct render) and $8$th
column (composite render)).

\newcommand{\cameraFigWidth}{0.111\textwidth}

\begin{figure*}
  \footnotesize
\renewcommand{\arraystretch}{0}
\addtolength{\tabcolsep}{-5.5pt}
\begin{tabular}{p{0.33cm}ccc|ccc|ccc}
  \rotatebox{90}{\ \quad Reference} &
  \includegraphics[width=\cameraFigWidth]{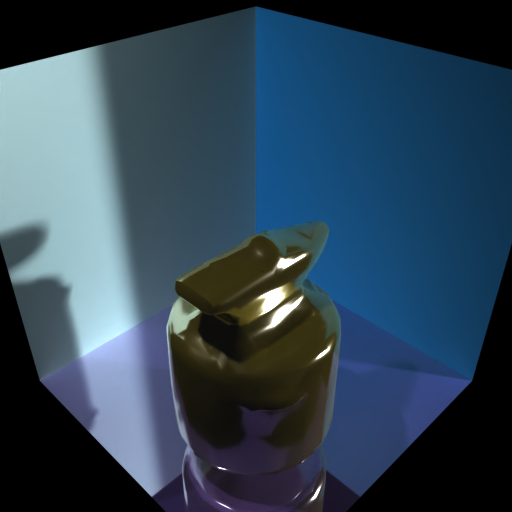} &
  \includegraphics[width=\cameraFigWidth]{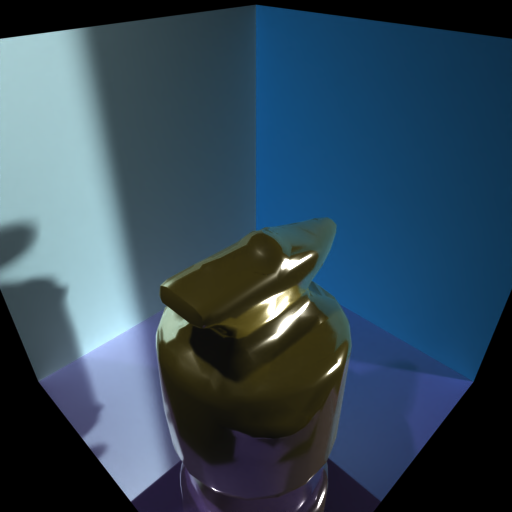} &
  \includegraphics[width=\cameraFigWidth]{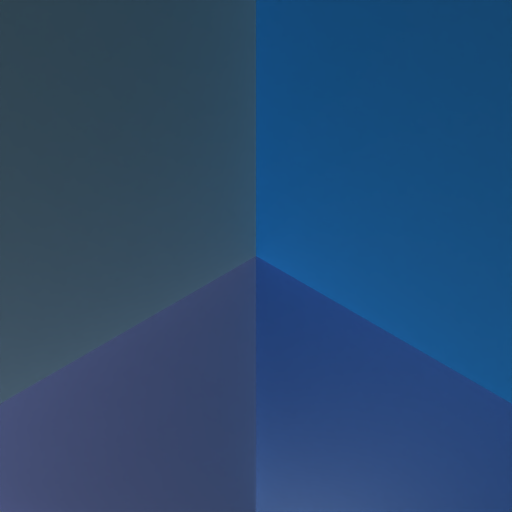} &
  \includegraphics[width=\cameraFigWidth]{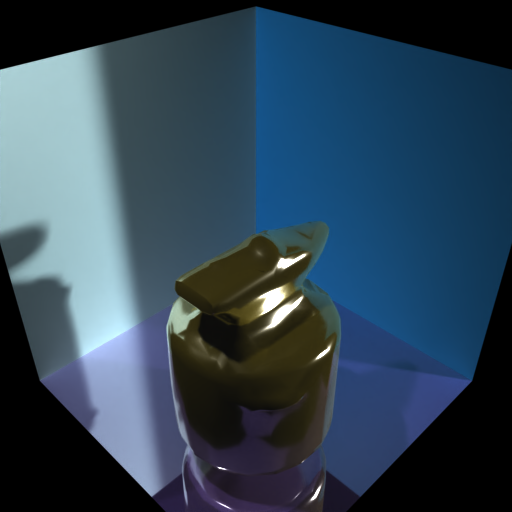} &
  \includegraphics[width=\cameraFigWidth]{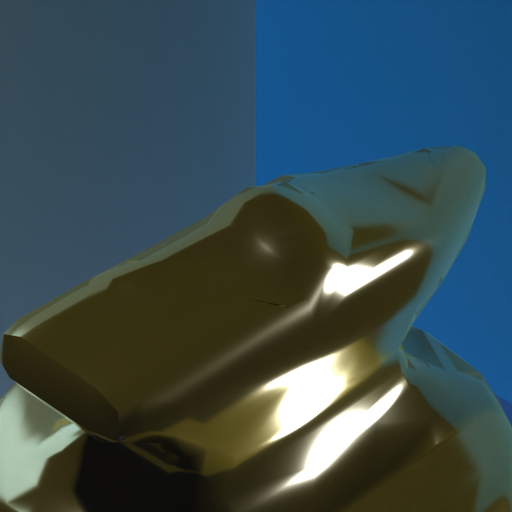} &
  \includegraphics[width=\cameraFigWidth]{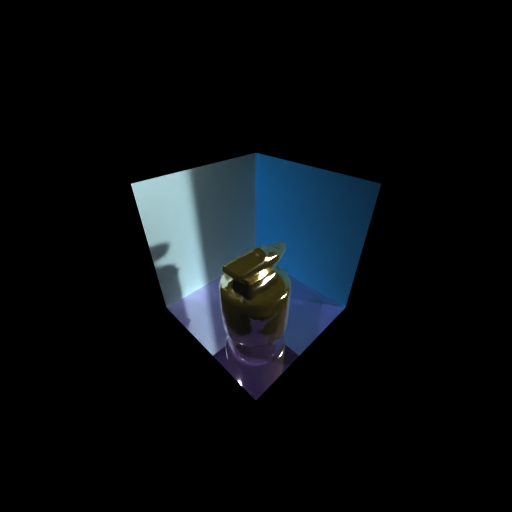} &
  \includegraphics[width=\cameraFigWidth]{figures/generalization/camera/train_dis_gt.png} &
  \includegraphics[width=\cameraFigWidth]{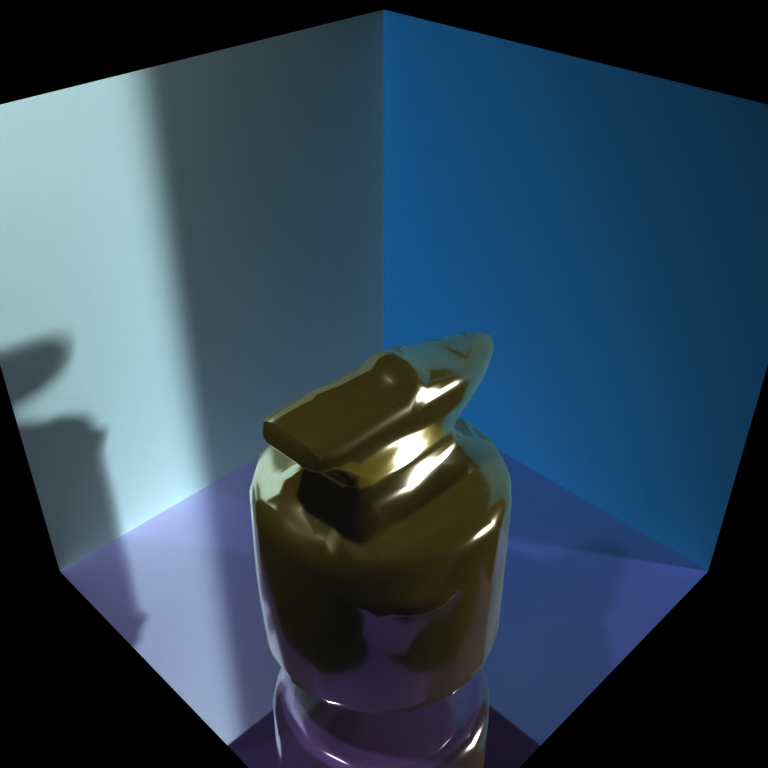} &
  \includegraphics[width=\cameraFigWidth]{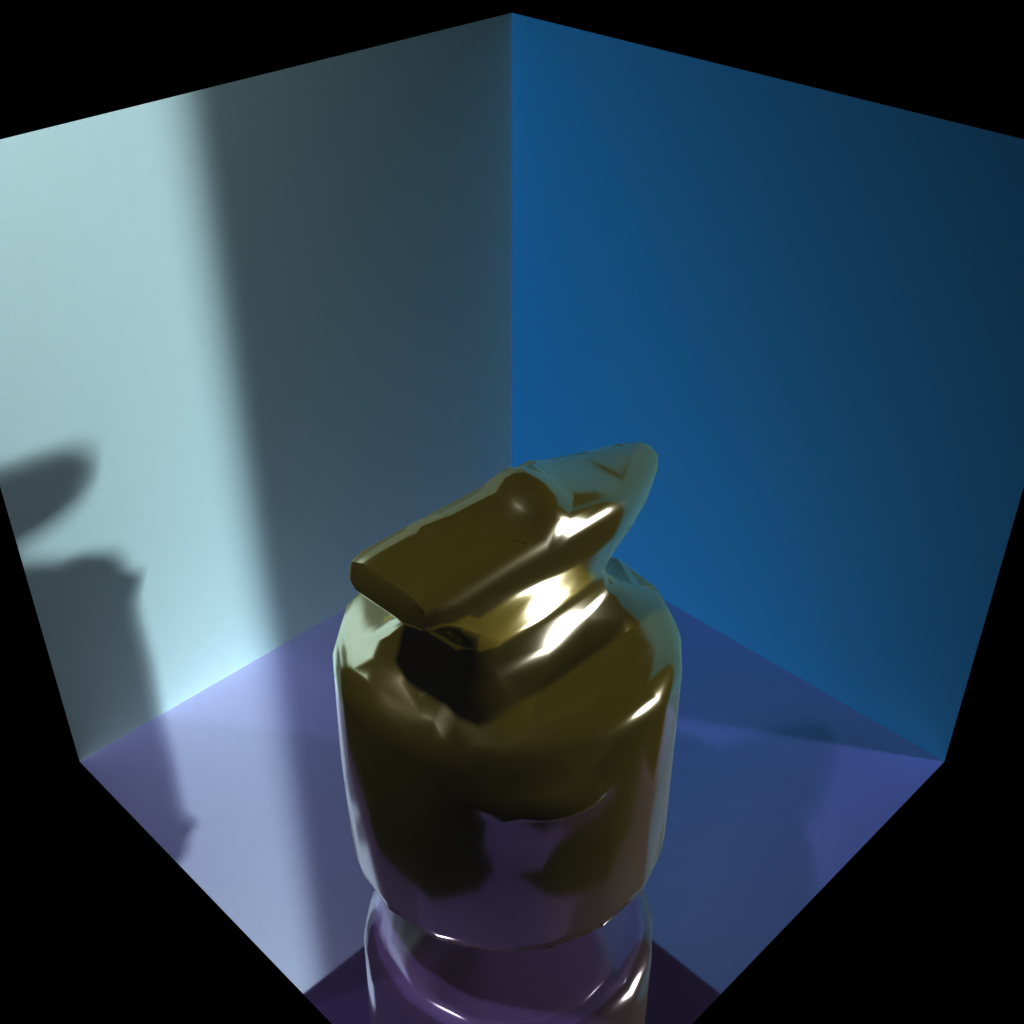} \\
  \rotatebox{90}{\ \ RenderFormer} &
  \includegraphics[width=\cameraFigWidth]{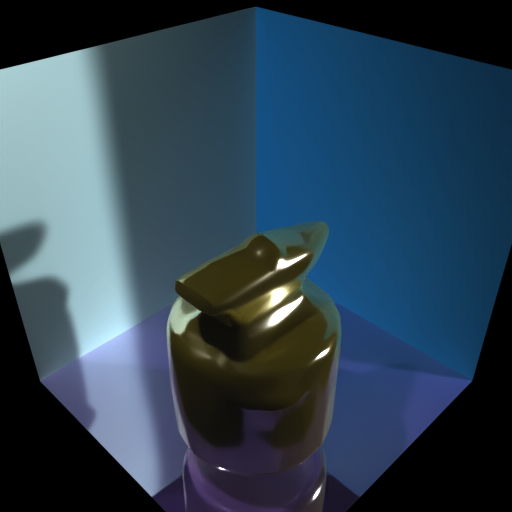} &
  \includegraphics[width=\cameraFigWidth]{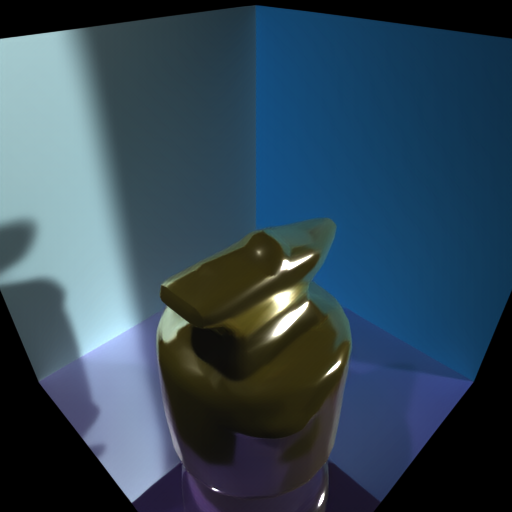} &
  \includegraphics[width=\cameraFigWidth]{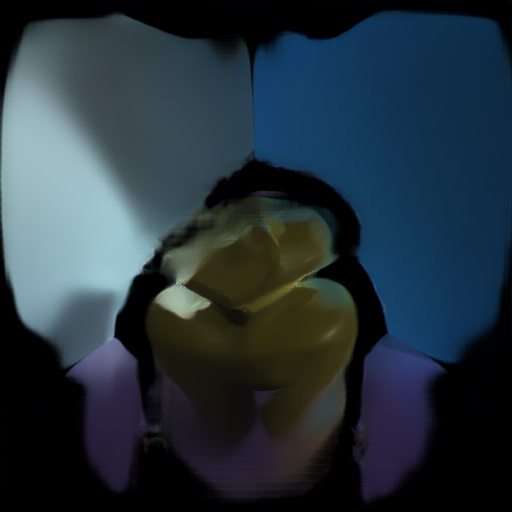} &
  \includegraphics[width=\cameraFigWidth]{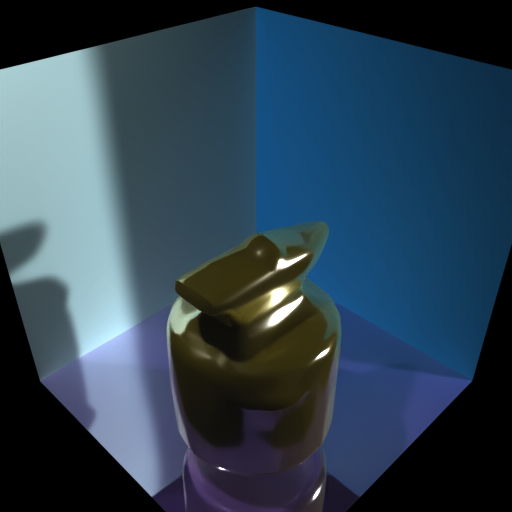} &
  \includegraphics[width=\cameraFigWidth]{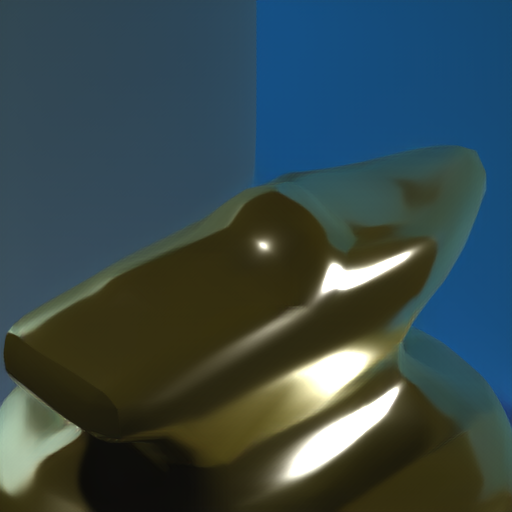} &
  \includegraphics[width=\cameraFigWidth]{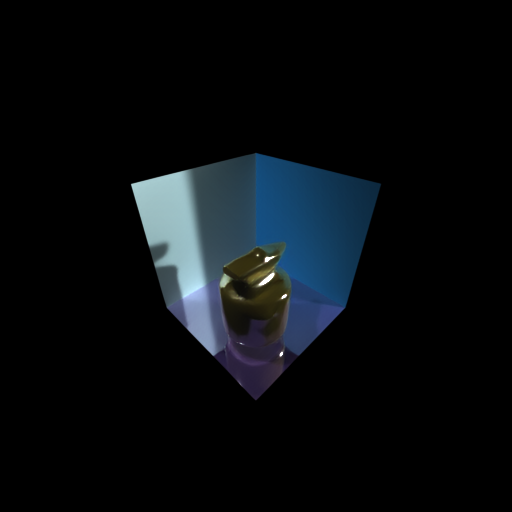} &
  \includegraphics[width=\cameraFigWidth]{figures/generalization/camera/train_dis_pred.png} &
  \includegraphics[width=\cameraFigWidth]{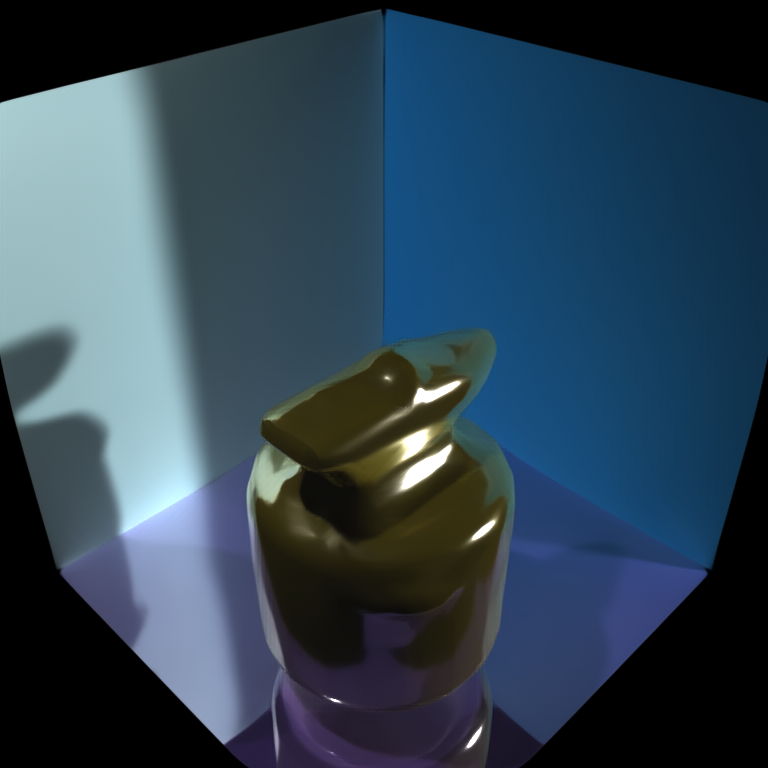} &
  \includegraphics[width=\cameraFigWidth]{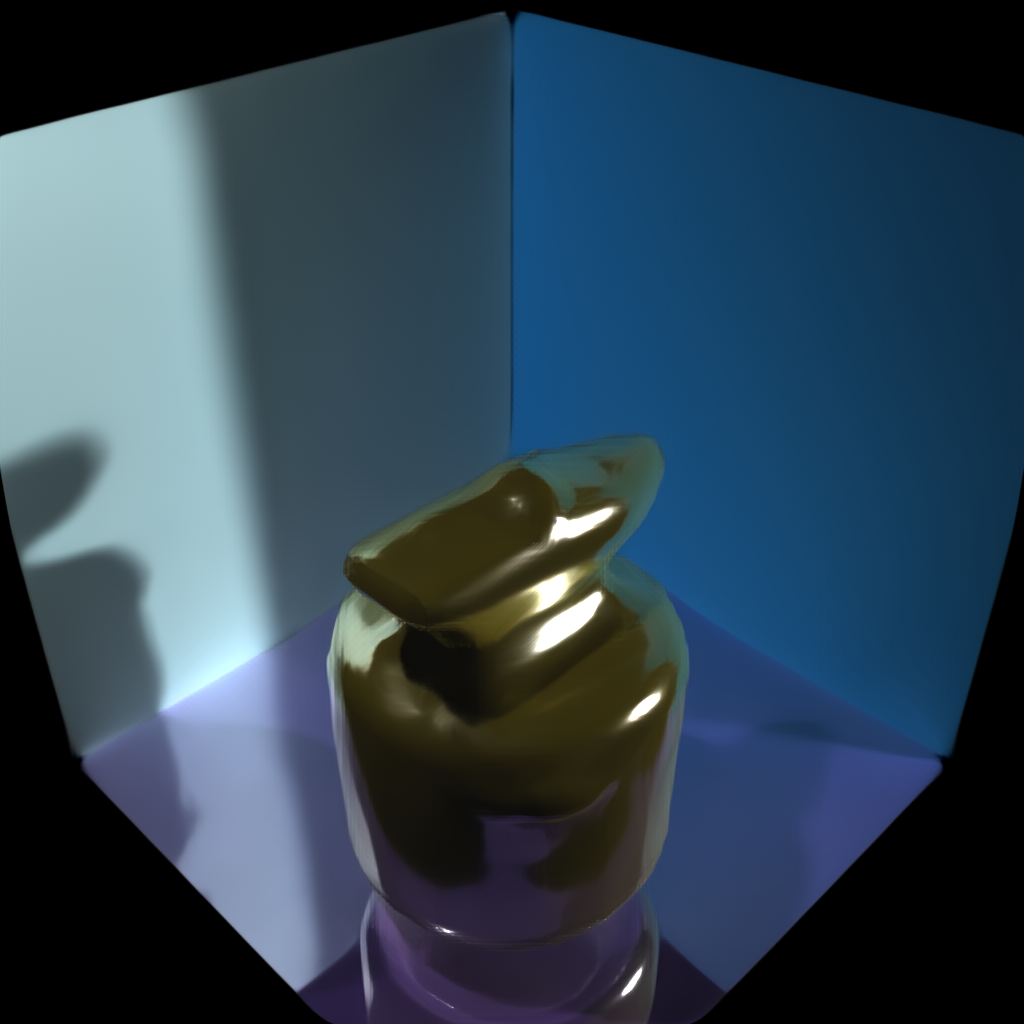} \\
  \vspace{0.1cm} \\
  & Normal & Close & Inside         &   Normal & Small & Large               &   $512 \times 512$ & $768 \times 768$ & $1024 \times 1024$ \\
  & \multicolumn{3}{c|}{Distance}   &   \multicolumn{3}{|c|}{Field of View}  &   \multicolumn{3}{|c}{Resolution} 
\end{tabular}
\caption{RenderFormer is robust to moving the camera closer than seen
  during training (2nd column), as long as the camera remains outside
  the scene (3rd column).  RenderFormer is also robust to exceeding
  the field of view seen during training (4th-6th column).  We also
  found that RenderFormer fails gracefully when rendering at higher
  resolutions (7th-9th column), with differences around depth
  discontinuities (\eg, between the gray and blue walls).}
  \label{fig:camera}
\end{figure*}

\newcommand{\shadowFigWidth}{0.1685\textwidth}
\begin{figure*}
\renewcommand{\arraystretch}{0.0}
\addtolength{\tabcolsep}{-5.5pt}
\begin{tabular}{ cccccc }
 \includegraphics[width=\shadowFigWidth]{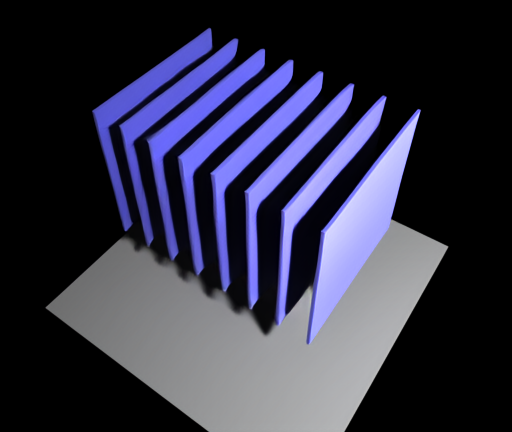}
 &\includegraphics[width=\shadowFigWidth]{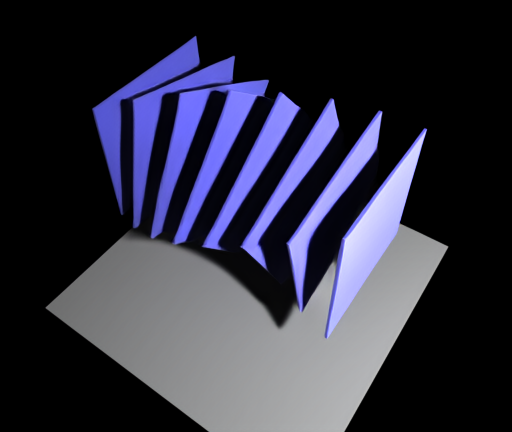}
 &\includegraphics[width=\shadowFigWidth]{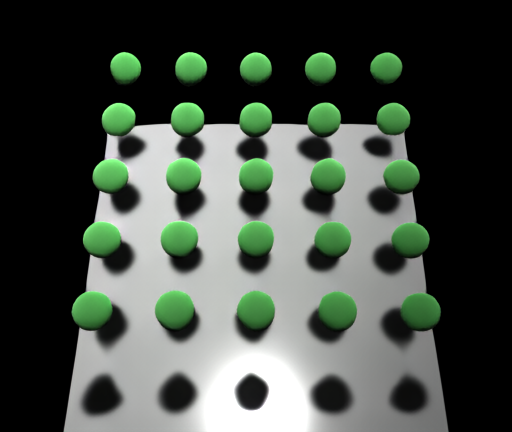}
 &\includegraphics[width=\shadowFigWidth]{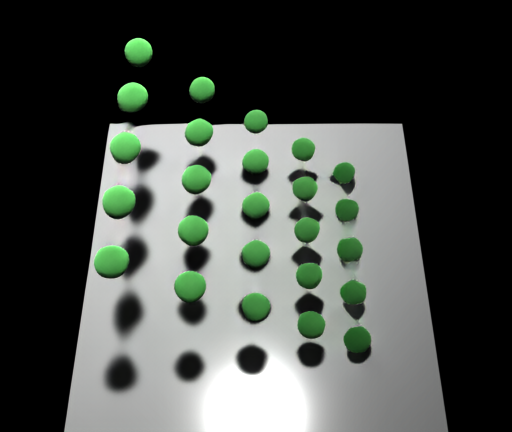}
 &\includegraphics[width=\shadowFigWidth]{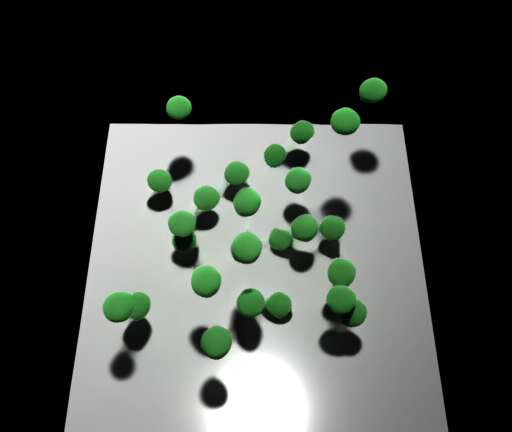}
 &\includegraphics[width=\shadowFigWidth]{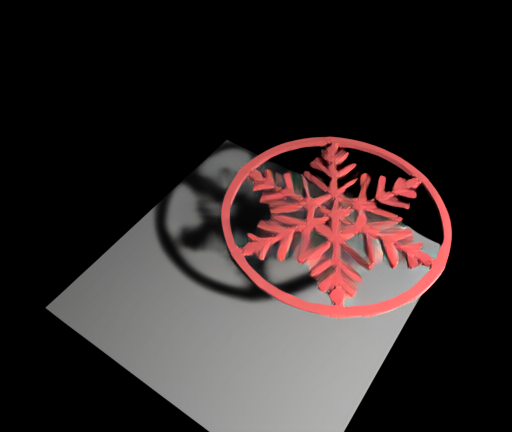} \\
\end{tabular}
\caption{RenderFormer can correctly reproduce occlusions for scenes
  with many objects.  However, the shadows cast by occluders with very
  complex shapes, can result in a loss of detail in the cast shadow.}
\label{fig:shadow}
\end{figure*}

\paragraph{Camera Parameters}
Similar to light sources, RenderFormer is also only trained for a
camera located outside the scene.  As a result RenderFormer has never
learned that triangles can be placed behind the camera, and fails to
correctly render such scenes (\autoref{fig:camera}, $3$nd column).
Extending the training set to include such cases could allow
RenderFormer to learn how to handle such cases.

RenderFormer is also trained for a limited range of FOV. However, from
\autoref{fig:camera} ($4$th to $6$th column) we can see that
RenderFormer appears to be robust to going outside this range.

RenderFormer is also only trained for a fixed $512 \times 512$
resolution.  While in theory the ray-bundles can model higher
resolution, we found that RenderFormer exhibits a minor resolution
dependence. As shown in~\autoref{fig:camera} ($8$th and $9$th column),
when applied to higher resolutions, RenderFormer fails gracefully,
with most of the errors focused around depth discontinuities.  We also
exploit this property by first training RenderFormer at lower
resolutions ($256 \times 256$), and then fine-tuning it at
$512 \times 512$ resolution.

\begin{figure*}
  {
    \footnotesize
    \newcommand{\specularFigWidth}{0.1685\textwidth}
    \renewcommand{\arraystretch}{0.0}
    \addtolength{\tabcolsep}{-5.5pt}
    \begin{tabular}{cc|cc|cc}
      \includegraphics[width=\specularFigWidth]{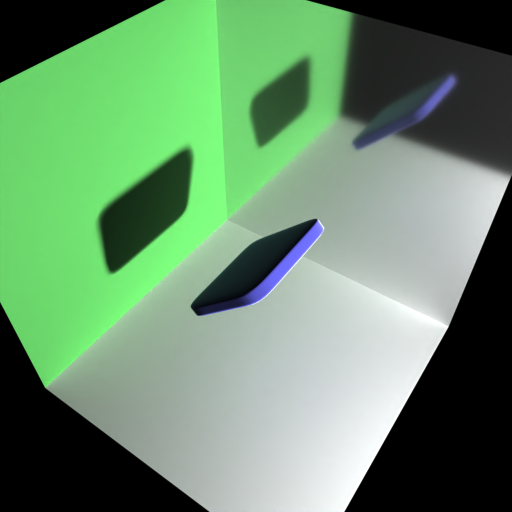} &
      \includegraphics[width=\specularFigWidth]{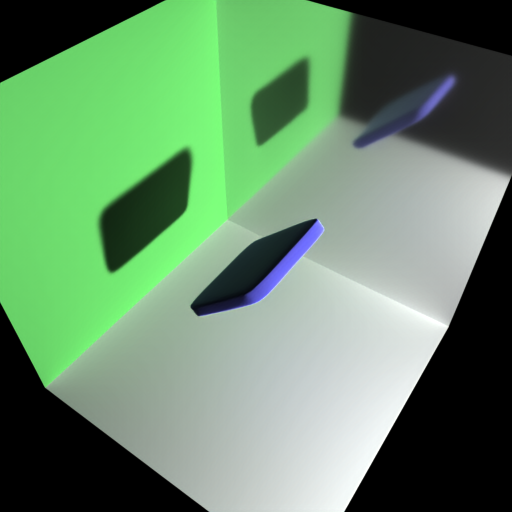} &
      \includegraphics[width=\specularFigWidth]{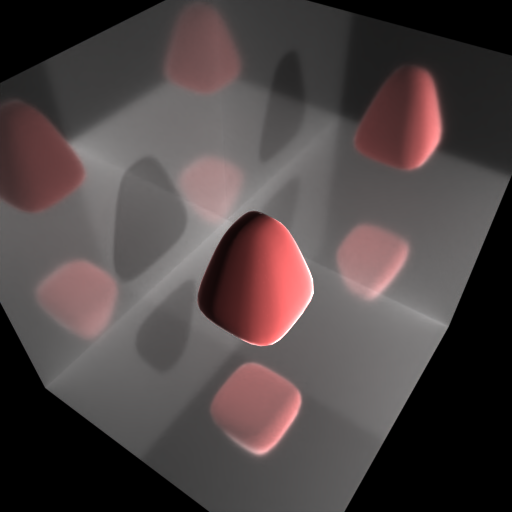} &
      \includegraphics[width=\specularFigWidth]{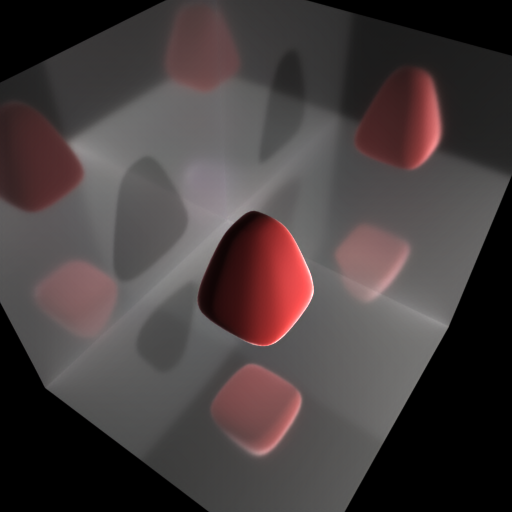} &
      \includegraphics[width=\specularFigWidth]{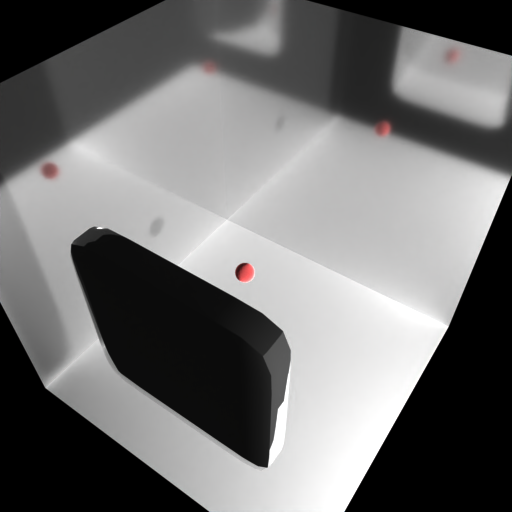} &
      \includegraphics[width=\specularFigWidth]{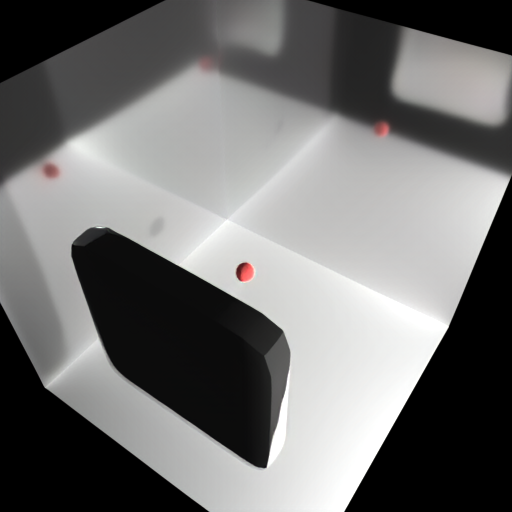} \\
    \vspace{0.1cm} \\
    Reference & RenderFormer & Reference & RenderFormer & Reference & RenderFormer \\
    \multicolumn{2}{c|}{1 Interreflection} & \multicolumn{2}{|c|}{2 Interreflections} & \multicolumn{2}{|c}{3 Interreflections}
    \end{tabular}
  }
    \caption{RenderFormer correctly handles $1$ and $2$ recursive
      specular interreflections. However, due to the scarcity of
      training exemplars with more specular interreflections, it does
      not always correctly resolve higher order reflections (\eg, the
      reflection of the red ball in the reflection of the mirror on
      the top wall).}
\label{fig:specular}
\end{figure*}
 
\begin{figure}
  \newcommand{\textureExtFigWidth}{0.16\textwidth}
  \renewcommand{\arraystretch}{0.0}
  \addtolength{\tabcolsep}{-5.5pt}
  \begin{tabular}{ ccc }
    \includegraphics[width=\textureExtFigWidth]{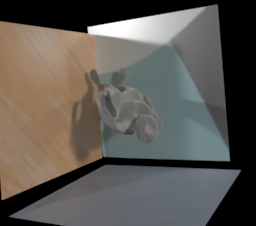}
    &\includegraphics[width=\textureExtFigWidth]{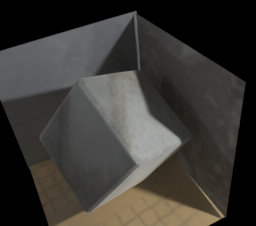} 
    &\includegraphics[width=\textureExtFigWidth]{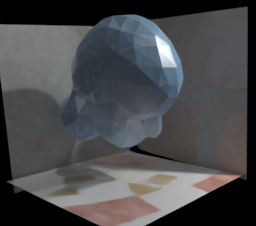}\\
  \end{tabular}
  \caption{Preliminary results of extending RenderFormer to support
    spatially-varying material properties.}
  \label{fig:textures}
\end{figure}

\paragraph{Scene Complexity}
Finally, we explore the accuracy of RenderFormer with respect to scene
complexity.  In particular, we investigate if RenderFormer can handle
complex occluders and multiple specular reflections.
\autoref{fig:shadow} shows scenes with increasing occluder
complexity. The first two columns show a series of closely packed
planes that cast shadows between the planes and the
floor. \autoref{fig:shadow} ($3$rd to $5$th column) shows a series of
small occluders placed at varying distances between the ground plane
and light source. We observe that the shadows of occluders with
complex shapes sometimes miss fine details (\autoref{fig:shadow},
columns $3$ to $5$).

\autoref{fig:specular} shows how well RenderFormer handles multiple
bounces of reflections. While RenderFormer does not reason in terms of
physical bounces of light transport, we find that RenderFormer
correctly models on average $3$ bounces of specular reflections, but
higher-order bounces are dropped (\eg, the 2nd reflection of red ball
on the back wall in the last example in~\autoref{fig:specular}).  We
found that the reflection depth is independent of the number of
view-dependent layers, and we posit that this limitation is mainly due
to the scarcity of training examples with higher-order specular
bounces, and careful augmentation of the training dataset could
improve performance for multi-bounce reflections.

\paragraph{Textures}
RenderFormer assumes constant reflectance properties over a triangle.
We perform an exploratory experiment to extend RenderFormer to include
spatially-varying surface reflectance by modifying the reflectance
token embedding.  Instead of expanding the stacked reflectance
parameters to a $768$-dimensional vector, we directly encode spatially
varying information at the triangle level. To embed the
spatially-varying parameters (\ie, $13$ channels containing diffuse
albedo, specular albedo, roughness, and surface normal), we first
rasterize the parameters to an isosceles right triangle at
$32 \times 32$ resolution. Next, we concatenate all texels in a
$13,\!312$-dimensional vector (\ie, $32 \times 32$ texels and $13$
channels per texel) that is encoded by a single linear layer followed
by RMS-Normalization into a $768$-dimensional token. Our initial
results (\autoref{fig:textures}) show that RenderFormer is able to
model spatially varying reflectance, albeit blurred.  Expanding the
token length might improve texture quality; we leave this for future
research.


\section{Conclusion}

In this paper we introduce RenderFormer, a transformer-based neural
rendering pipeline that takes as input a regular triangle mesh, and
outputs an image of the scene accounting for global illumination.
While RenderFormer is limited in the scenes it can render (\ie,
limited triangle count, number of light sources, camera positions,
etc...), it generalizes better than prior neural rendering systems.
RenderFormer approaches solving light transport in a virtual scene as
a two-stage sequence-to-sequence transformation. The first stage
transforms a triangle-sequence to model view-independent
triangle-to-triangle transport. The second stage transforms a sequence
of ray-bundles to a sequence of corresponding observed radiance values
guided by the triangle-sequence from the first stage.

There are ample avenues to further improve RenderFormer.  First, we
can expand the training set to support a wider variety of camera and
light positions. Furthermore, while our current implementation
utilizes training data rendered using the GGX BRDF model, we impose no
inherent architectural restrictions related with the reflectance
model. Hence, RenderFormer could be trained on alternative datasets
using other reflectance models including those that model transparency
or subsurface scattering.  Currently, RenderFormer only supports
simple light sources, and extensions to environment lighting and
non-diffuse light sources would further generalize RenderFormer.
Since RenderFormer is fully transformer based, it is inherently
differentiable, allowing us to train RenderFormer directly from
data. An interesting and promising direction for future work that
leverages the inherent differentiability, would be to apply
RenderFormer to inverse rendering applications.  Finally, would like
to investigate hierarchical attention methods based on existing
grouping based acceleration structures for classic rendering methods
(e.g., BVH) to support more complex scenes with larger
triangle-meshes.

\begin{acks}
  We would like to thank Kexun Zhang and Kaiqi Chen for discussions on
  transformer model design and performance optimizations, and Sam
  Sartor for Blender Cycle tips and pre-reviewing this work.  Pieter
  Peers was supported in part by NSF grant IIS-1909028. Chong Zeng and
  Hongzhi Wu were partially supported by NSF China (62332015, 62227806
  \& 62421003), the XPLORER PRIZE, and Information Technology Center
  and State Key Lab of CAD\&CG, Zhejiang University.
\end{acks}

\bibliographystyle{ACM-Reference-Format}
\bibliography{src/reference}

\end{document}